\documentclass[11pt,xcolor=dvipsnames,fleqn]{article}
\DeclareMathAlphabet{\scr}{U}{rsfs}{m}{n}

\pdfoutput=1
\usepackage{scalerel,stackengine}
\usepackage[colorlinks,urlcolor=black,linkcolor = blue,citecolor = black,]{hyperref}
\usepackage{subfigure}
\usepackage{latexsym}
\usepackage{epsfig}
\usepackage[mathscr]{eucal}
\usepackage{amsfonts}
\usepackage{amscd}
\usepackage{cite}
\usepackage{array}
\usepackage{amssymb}
\usepackage{colordvi}
\usepackage[centertags]{amsmath}
\usepackage{enumerate}
\usepackage{graphicx}
\usepackage{booktabs}
\usepackage{theorem}
\usepackage[footnotesize]{caption}
\usepackage{soul}
\usepackage{mcite}
\usepackage{slashed}
\usepackage{braket}
\usepackage{xcolor}
\usepackage{bbm}
\usepackage[utf8]{inputenc}
\usepackage{fancyvrb}
\usepackage{framed}
\usepackage{xspace}
\usepackage{todonotes}
\usepackage{tikz-feynman}
\usepackage{siunitx}

\newcommand{\be}{\begin{equation}}
\newcommand{\ee}{\end{equation}}
\newcommand{\bea}{\begin{eqnarray}}
\newcommand{\eea}{\end{eqnarray}}

\newcommand{\beq}{\begin{eqnarray}}
\newcommand{\eeq}{\end{eqnarray}}
\newcommand{\bpmatrix}{\begin{pmatrix}}
\newcommand{\epmatrix}{\end{pmatrix}}
\newcommand{\ba}{\begin{array}}
\newcommand{\ea}{\end{array}}

\newcommand{\al}{\alpha}

\renewcommand{\Re}{\text{Re}\!}

\newcommand{\OS}{\text{OS}}
\newcommand{\MSb}{\overline{\text{MS}}}

\newcommand{\eg}{{\it e.g.\;}}
\newcommand{\bc}{\begin{center}}
\newcommand{\ec}{\end{center}}

\newcommand{\gev}{~\text{GeV}}

\newcommand{\tev}{~\text{TeV}}

\newcommand{\ii}{\mathit{i}}

\clearpage{\pagestyle{empty}\cleardoublepage}

\newcommand{\cbrak}[1]{\left(#1\right)}
\newcommand{\sbrak}[1]{\left[#1\right]}
\newcommand{\vS}{v_s}
\newcommand{\X}{\chi}
\newcommand{\mX}{m_{\chi}}

\newcommand{\mh}{m_{h_1}}
\newcommand{\mH}{m_{h_2}}
\newcommand{\GG}{G_{\mu\nu}^aG^{a\mu\nu}}
\newcommand{\upV}{\textit{upV}}
\newcommand{\loV}{\textit{loV}}
\newcommand{\model}{PNGDM}
\definecolor{blobColor}{RGB}{191,191,191}
\allowdisplaybreaks

\usepackage{cleveref}
\crefname{chapter}{Chapter}{Chapter}
\crefname{section}{Sec.}{Secs.}
\crefname{table}{Tab.}{Tabs.}
\crefname{figure}{Fig.}{Figs.}
\crefname{equation}{Eq.}{Eqs.}
\crefname{appendix}{Appendix\ }{Appendix\ }

\begin{document}
\pdfoutput=1

\title{
    \vspace*{-3.7cm}
    \phantom{h} \hfill\mbox{\small KA-TP-12-2020}\\[-1.1cm]
    \vspace*{2.7cm}
    \textbf{Electroweak Corrections in a Pseudo-Nambu Goldstone Dark Matter Model Revisited \\[4mm]}}

\date{}
\author{
Seraina Glaus$^{1,2\,}$\footnote{E-mail:
    \texttt{seraina.glaus@kit.edu}} ,
Margarete M\"{u}hlleitner$^{1\,}$\footnote{E-mail:
    \texttt{milada.muehlleitner@kit.edu}} ,
Jonas M\"{u}ller$^{1\,}$\footnote{E-mail:
    \texttt{jonas.mueller@kit.edu}} ,
\\
Shruti Patel$^{1,2\,}$\footnote{E-mail: \texttt{shruti.patel@kit.edu}},
Tizian R\"{o}mer$^{1\,}$\footnote{{E-mail: \texttt{tizian.roemer@student.kit.edu}}},
Rui Santos$^{3,4\,}$\footnote{E-mail:
    \texttt{rasantos@fc.ul.pt}}
\\[5mm]
{\small\it
$^1$Institute for Theoretical Physics, Karlsruhe Institute of Technology,} \\
{\small\it 76128 Karlsruhe, Germany}\\[3mm]
{\small\it
$^2$Institute for Nuclear Physics, Karlsruhe Institute of Technology,
76344 Karlsruhe, Germany}\\[3mm]
{\small\it $^3$Centro de F\'{\i}sica Te\'{o}rica e Computacional,
Faculdade de Ci\^{e}ncias,} \\
{\small \it    Universidade de Lisboa, Campo Grande, Edif\'{\i}cio C8
1749-016 Lisboa, Portugal} \\[3mm]
{\small\it
$^4$ISEL -
Instituto Superior de Engenharia de Lisboa,} \\
{\small \it   Instituto Polit\'ecnico de Lisboa
1959-007 Lisboa, Portugal} \\[3mm]
}

\maketitle

\begin{abstract}
Having so far only indirect evidence for the existence of Dark Matter
a plethora of experiments aims at direct detection of Dark Matter
through the scattering of Dark Matter particles off atomic nuclei. For
the correct interpretation and identification of the underlying nature
of the Dark Matter constituents higher-order corrections to the cross
section of Dark Matter-nucleon scattering are important, in particular
in models where the tree-level cross section is negligibly small.  In
this work we revisit the electroweak corrections to the dark
matter-nucleon scattering cross section in a model with a pseudo
Nambu-Goldstone boson as the Dark Matter candidate. Two calculations
that already exist in the literature, apply different approaches
resulting in different final results for the cross section in some regions of
the parameter space leading us to redo the calculation and analyse
the two approaches to clarify the situation. We furthermore update the
experimental constraints and examine the regions of the parameter
space where the cross section is above the neutrino floor but which
can only be probed in the far future. 
\end{abstract}
\thispagestyle{empty}
\vfill
\newpage
\setcounter{page}{1}
\section{Introduction} 
Ever since Dark Matter (DM) became an inevitable ingredient in model building, all kind of proposals integrating DM candidates into phenomenologically viable models
have emerged, from the most simple extensions of the Standard Model (SM) to 
fairly intricate models. The accumulated data from astrophysics and cosmology strongly suggests that if DM is a particle, it is most
probably cold and with a mass close to the electroweak scale. Particles with these features are known as Weakly Interacting Massive Particles (WIMPs). 
In this work we study a simple extension of the SM where a complex singlet is added to the SM field content. The model
is built such that after spontaneous symmetry breaking one of the
singlet components will mix with the SM-like Higgs boson while the other one will play the role of the DM candidate. 
This type of extension was first proposed in Refs.~\cite{Silveira:1985rk,
  McDonald:1993ex, Burgess:2000yq, Bento:2000ah} 
and recently reviewed with updated experimental constraints in Ref.~\cite{Azevedo:2018oxv}.

A particular version of this extension known as the Pseudo Nambu-Goldstone DM model (\model) has a scalar potential invariant under a global $U(1)$ symmetry
which would give rise to a Nambu-Goldstone boson. The symmetry is then softly broken and a pseudo Nambu-Goldstone boson emerges as the Dark Matter candidate.
As discussed in previous works~\cite{PhysRevLett.119.191801,
  Azevedo:2018exj}, the nature of this particle makes the DM-nucleon
tree-level cross section proportional to the velocity of the DM
particle and therefore negligible 
(see also Ref.~\cite{Burgess:1998ku} for an interesting discussion on the subject of Goldstone and pseudo-Goldstone bosons). Hence,
the leading order cross section is given by its one-loop contribution. The first calculation of the electroweak corrections to this process was performed
in Ref.~\cite{Azevedo:2018exj} and almost at the same time a second calculation appeared in Ref.~\cite{Ishiwata:2018sdi}. In this
work we will perform once again the calculation of the electroweak corrections to the DM-nucleon cross section while discussing in detail the main 
differences with respect to the two previous
calculations and the reasons for settling this issue. Our calculation will be performed with
a different renormalisation scheme. This allows us to
perform a rough estimate of the remaining uncertainties on the cross
section due to missing higher-order corrections. 

We then perform a scan in the parameter space taking into account the most relevant theoretical and experimental constraints. We will show that there
are still allowed points in the parameter space above the neutrino floor~\cite{Billard:2013qya}
 but only experiments in the far future will be able to probe them.       

The paper is organized as follows. In section~\ref{sec:model} we
briefly present the complex singlet extension of the SM while in
section~\ref{sec:renmodel} we introduce the renormalisation 
of the model. The various aspects of the DM direct-detection cross section at tree level and at one-loop level are discussed in~\ref{sec:xsection}.  
In section~\ref{sec:Res} we discuss the results and future prospects of DM detection in this model taking into account the most recent
constraints. We summarise our findings in section~\ref{sec:Conc}.

\section{The Model}
\label{sec:model}

A simple extension of the SM by a scalar gauge singlet is enough to provide a valid DM candidate. 
The new complex scalar field $S$, with zero hypercharge and zero
isospin only enters the model via the scalar potential that can be written as 
\begin{equation}
    V = - \frac{\mu_H^2}{2} \left|H\right|^2 + \frac{\lambda_H}{2} \left|H\right|^4 - \frac{\mu_S^2}{2} \left|S\right|^2 + \frac{\lambda_S}{2} \left|S\right|^4 + \lambda_{HS} \left|H\right|^2 \left|S\right|^2 - \frac{\mX^2}{4} (S^2 + S^{*2})\,,
    \label{eq:potential}
\end{equation}
where the mass parameters $\mu_H^2$, $\mu_S^2$, $\mX^2$ and the quartic couplings $\lambda_S$, $\lambda_H$, $\lambda_{HS}$ are real due to hermicity. The doublet $H$ and singlet $S$ fields are expanded as follows
\begin{equation}
    H = \begin{pmatrix}
        G^+ \\\frac{1}{\sqrt{2}} \left( v + \Phi_H + i G^0 \right)
    \end{pmatrix}
    \,, \qquad S = \frac{1}{\sqrt{2}} \left( \vS + \Phi_S + i\X \right) \, ,\label{eq:vevsExpansion}
\end{equation}
with the electroweak vacuum expectation value (VEV) $v$ and the singlet VEV $\vS$. 
With this definition the model is invariant under the DM charge conjugation $S\rightarrow S^*$, which guarantees the stability of the imaginary part of $S$. 
Furthermore, in order to simplify the potential, an invariance under
the $Z_2$ symmetry $S \to - S$ has also been imposed.
The real part of $S$ develops a vacuum expectation value (VEV), while the doublet develops the usual (SM) VEV that gives mass to the SM fermions and gauge
bosons, 
\begin{equation}
    \braket{H} = \frac{1}{\sqrt{2}}\begin{pmatrix}
        0 \\v
    \end{pmatrix}
    \,,
    \quad
    \braket{S}=\frac{\vS}{\sqrt{2}}
\end{equation}
with $v=2 m_W/g$, $m_W$ the $W$ boson mass and $g$ the $SU(2)$ coupling constant. 
Because the real part of $S$ acquires a VEV, $\Phi_S$ cannot be a viable DM candidate and
it mixes with the doublet real neutral component $\Phi_H$. Using the minimum conditions
\begin{equation}
    \begin{split}
        \left. \frac{\partial V}{\partial H} \right|_\text{VEV} &= 0 \qquad \Longleftrightarrow \qquad  T_H \equiv \frac{v}{2}  \left(-\mu_H^2 + \lambda_H v^2 + \lambda_{HS} \vS^2\right) = 0\,,\\
        \left. \frac{\partial V}{\partial S} \right|_\text{VEV} &= 0
        \qquad \Longleftrightarrow \qquad T_S \equiv \frac{\vS}{2}
        \left(-\mu_S^2 + \lambda_S \vS^2 +
          \lambda_{HS} v^2 - \mX^2\right) = 0\,.
    \end{split}\label{eq:miniCon}
\end{equation}
we can write the mass matrix of the two neutral states as 
\begin{equation}
    \mathcal{M}_T^2 = \mathcal{M}^2 + \mathcal{T}\,, \quad \mathcal{M}^2 =
    \begin{pmatrix}
        \lambda_H v^2     & \lambda_{HS} v \vS \\
        \lambda_{HS} v \vS & \lambda_{S} \vS^2
    \end{pmatrix}, \quad
    \mathcal{T} = \begin{pmatrix}
        T_H/v & 0 \\0&T_S/\vS
    \end{pmatrix}, \label{eq:massMatrix}
\end{equation}
The mass eigenstates $h_1$ and $h_2$ are obtained from the gauge eigenstates via
\begin{equation}
   \begin{pmatrix}
        h_1 \\h_2
    \end{pmatrix} \equiv R(\al)
    \begin{pmatrix}
        \Phi_H \\\Phi_S
    \end{pmatrix}, \label{eq:diagM}
\qquad
    M^2 \equiv R(\al) \, \mathcal{M}^2 \, R^{-1}(\al) =
    \begin{pmatrix}
        \mh^2 & 0 \\0&\mH^2
    \end{pmatrix}
\end{equation}
with the orthogonal matrix $R(\alpha)$
\begin{equation}
    R(\al) \equiv
    \begin{pmatrix}
        \cos\alpha & \sin\alpha \\-\sin\alpha&\cos\alpha \, 
    \end{pmatrix} \, . \label{eq:matrixR}
\end{equation}
One of these mass eigenstates is identified as the 125 GeV Higgs
boson. 
The DM particle is given by $\chi$. Exploiting the
  tadpole conditions Eq.~(\ref{eq:miniCon}) its mass can be written as
\beq
m_\chi^2 + \frac{T_S}{v_S} \;.
\eeq
We also require that the potential is bounded from below inducing the tree-level conditions
\begin{equation}
    \lambda_H > 0, \qquad \lambda_S>0, \qquad \lambda_{HS} > -\sqrt{\lambda_H\lambda_S}\,.
\end{equation}
The parameters of the potential can be written as functions of the masses, mixing angle and the VEVs as
\begin{equation}
    \begin{split}
        \lambda_{HS} &= -\frac{\mH^2 - \mh^2}{2v\vS}\sin 2\alpha\,,\\
        \lambda_H &= \frac{\mH^2 \sin^2 \alpha + \mh^2 \cos^2 \alpha}{v^2}\,,\\
        \lambda_S &= \frac{\mH^2 \cos^2 \alpha + \mh^2 \sin^2 \alpha}{\vS^2}\,. \label{eq:lambdas}
    \end{split}
\end{equation}
 We choose the following parameters as
  independent input parameters,
\begin{equation}
    v\,, \, \, \vS\,, \, \, \alpha\,, \, \, \mh^2\,, \, \, \mH^2\,, \, \, \mX^2\,, \, \, T_H\,, \, \, T_S\,.\label{eq::InputParameter}
\end{equation}

Some final comments regarding the scalar potential are
in order. The potential is invariant under a $U(1)$ symmetry
($S \to e^{i\alpha} S$) that is softly broken by the dimension-two term proportional to $m_\chi^2$.
The Goldstone boson related to the $U(1)$ symmetry acquires a mass
proportional to $m_\chi^2$. Due to the $Z_2$
symmetry there are no more terms contributing to the mass of the 
pseudo Nambu-Goldstone boson. Hence, the $U(1)$ symmetry is recovered by setting $m_\chi^2 =0$, where the true Nambu-Goldstone
boson is recovered.

\section{Renormalisation of the PNGDM}
\label{sec:renmodel}

In the following, we present the renormalisation of the \model{} in
order to be able to calculate the 
electroweak (EW) corrections to the scattering process of the
pseudo-Nambu Goldstone DM particle with a nucleon. Having defined the
full set of input parameters in \cref{eq::InputParameter}, the bare
parameters $p_0$ are replaced with the renormalized
ones, $p$, according to
\begin{equation}
    p_0 = p +\delta p\,,
\end{equation}
where $\delta p$ corresponds to the counterterm of the respective bare
parameter $p_0$. For a generic bare
field $\Psi_0$ (scalar, fermion or vector
field), the renormalized field $\Psi$ is expressed as
\begin{equation}
    \Psi_0 = \sqrt{Z_{\Psi}} \Psi\,,
\end{equation}
with the field strength renormalisation constant
$Z_{\Psi}\equiv 1 + \delta Z_\Psi$. Note that
$Z_{\Psi}$ is a matrix for mixing fields as present in
the \model. 
Dropping for simplicity the index 0 in the following,
  we hence make the following replacements
\beq
p &\to& p + \delta p \\
\Psi &\to& \left( 1 + \frac{1}{2} \delta Z_\Psi \right) \Psi \;,
\eeq
and analogously for the tadpole parameter
\beq
T \to T + \delta T \;.
\eeq
In the following we discuss each sector separately.

\subsection{Gauge Sector}
\label{sec::GaugeSector}
The gauge sector of the \model{} is not extended compared to the
SM. To set our notation and conventions we will list the counterterms
of the gauge sector in the following. We choose to
perform the renormalisation in the
mass basis of the \model, so that the following set of on-shell (\OS)
counterterms are taken for the gauge sector 
\begin{subequations}
    \begin{eqnarray}
        m_W^2 \rightarrow m_W^2 + \delta m_W^2\,,\\
        m_Z^2 \rightarrow m_Z^2 + \delta m_Z^2\,,\\
        e \rightarrow e + \delta Z_e e \,,\\
        g \rightarrow g + \delta g\,,
    \end{eqnarray}
\end{subequations}
As $m_W$ and $g$ have already been defined above: \\
where $m_Z$ is the mass of the EW neutral gauge boson $Z$ and the
electric coupling is denoted by $e$. The
renormalized fields are obtained through the
field strength renormalisation constants as
\begin{subequations}
    \begin{eqnarray}
        W^{\pm}\rightarrow \cbrak{1 + \frac{1}{2} \delta Z_{WW}} W^{\pm}\,,\\
        \begin{pmatrix}
            Z \\\gamma
        \end{pmatrix}
        \rightarrow
        \begin{pmatrix}
            1+\frac{1}{2} \delta Z_{ZZ}    & \frac{1}{2} \delta Z_{Z\gamma}          \\
            \frac{1}{2}\delta Z_{\gamma Z} & 1 + \frac{1}{2} \delta Z_{\gamma\gamma}
        \end{pmatrix}
        \begin{pmatrix}
            Z \\ \gamma
        \end{pmatrix}\,.
    \end{eqnarray}
\end{subequations}
Applying \OS{} conditions yields the following mass counterterms
\begin{equation}
    \delta m_W^2 = \Re \, \Sigma_{WW}^{T}\cbrak{m_W^2}\,\quad \text{and}\quad\delta m_W^2 = \Re \, \Sigma_{ZZ}^{T}\cbrak{m_Z^2}\,,
\end{equation}
with $T$ indicating the transverse part of the self-energies $\Sigma_{ii}~(ii=W,Z)$. 
The counterterm for the gauge coupling $g$, is obtained from the one for the electric charge and the one for the Weinberg angle $\theta_W$ using
\begin{eqnarray}
    e = g \sin \theta_W\,,\quad \text{with}\quad \cos\theta_W = \frac{m_W}{m_Z}\, .
\end{eqnarray}
The electric charge counterterm
is fixed in the Thomson limit, which by making use of Ward identities
allows us to write
\cite{Denner:1991kt}\footnote{Note
      that the sign in the second term of Eq.~(\ref{eq:deltae})
      differsfrom the one in \cite{Denner:1991kt} due to different sign
      conventions in the covariant derivative.}
\begin{eqnarray}
    \delta Z_e =
  \frac{1}{2}\left. \frac{\partial\Sigma_{\gamma\gamma}^T(p^2)}{\partial
  p^2}\right|_{p^2=0} + \frac{s_W}{c_W} \frac{\Sigma_{\gamma
  Z}^T(0)}{m_Z^2}\,, \label{eq:deltae}
\end{eqnarray}
where we introduced the short-hand notation $s_W
  \equiv \sin \theta_W$, $c_W\equiv \cos \theta_W$, and
\begin{eqnarray}
    \frac{\delta g}{g} = \delta Z_e + \frac{1}{2} \frac{1}{m_Z^2-m_W^2} \cbrak{\delta m_W^2 - c_W^2 \delta m_Z^2}\,.
\end{eqnarray}
The corresponding wave-function renormalisation constants guaranteeing
the correct \OS~properties are given by
\begin{subequations}
\begin{eqnarray}
    \delta Z_{WW} = -\Re \,\frac{\partial\Sigma_{WW}^2(p^2)}{\partial
        p^2}\bigg\vert_{p^2=m_W^2} \,,
    \\
    \begin{pmatrix}
        \delta Z_{ZZ}       & \delta Z_{Z\gamma}      \\
        \delta Z_{\gamma Z} & \delta Z_{\gamma\gamma}
    \end{pmatrix}
    =
    \begin{pmatrix}
        -\Re\, \frac{\partial \Sigma_{ZZ}^T(p^2)}{\partial
        p^2}\bigg\vert_{p^2=m_Z^2}             & 2
        \frac{\Sigma_{Z\gamma}^T(0)}{m_Z^2}                            \\
        -2 \frac{\Sigma_{Z\gamma}^T(0)}{m_Z^2} & -\Re \,\frac{\partial
            \Sigma_{\gamma\gamma}^T(p^2)}{\partial p^2}\bigg\vert_{p^2=0}
    \end{pmatrix}\,.
\end{eqnarray}
\end{subequations}

\subsection{Scalar Sector}

In the \model{} we have two additional scalars, one extra CP-even
Higgs boson and the DM candidate $\X$. The two CP-even scalars are mass-ordered as $h_1$ and $h_2$ with $m_{h_1}< m_{h_2}$ 
and the SM-like Higgs boson, with a mass of 125.09
  GeV, can be either of them. We again use an
on-shell scheme for the fields. The field strength renormalisation
constants read
\begin{equation}
    \begin{pmatrix}
        h_1 \\ h_2
    \end{pmatrix}
    \rightarrow
    \begin{pmatrix}
        1+\frac{1}{2}\delta Z_{h_1 h_1} & \frac{1}{2}\delta Z_{h_1 h_2}    \\
        \frac{1}{2} \delta Z_{h_2 h_1}  & 1+\frac{1}{2} \delta Z_{h_2 h_2}
    \end{pmatrix}
    \begin{pmatrix}
        h_1 \\ h_2
    \end{pmatrix}\,.
    \label{renorm::ZfactorHiggs}
\end{equation}
The mass matrix with the additional tadpole contributions is given by 
\begin{equation}
    \mathcal{M}_{h_1 h_2} =
    \begin{pmatrix}
        m_{h_1}^2 & 0         \\
        0         & m_{h_2}^2
    \end{pmatrix}
    +
    \underbrace{
        R(\al) \begin{pmatrix}
            T_{H}/v & 0              \\
            0            & T_{S}/v_S
        \end{pmatrix}
        R(\al)^T
    }_{\equiv\delta T}\,.
    \label{RENORM::MASSMATRIX}
\end{equation}
The rotation matrix $R(\al)$ is defined in \cref{eq:matrixR} and diagonalises the gauge eigenstates in the Higgs mass basis.
The tadpole terms $\delta T$ in the tree-level mass matrix are bare
parameters. At next-to-leading (NLO) they get shifted due to EW
corrections to the vaccuum state of the potential. Defining the
tree-level VEV to be the same to all orders of perturbation theory,
requires the introduction of tadpoles counterterms such that the the one-loop renormalized
one-point function $\hat{T}_i$ ($i=H,S$) vanishes
\begin{equation}
    \hat T_i =T_i - \delta T_i \overset{!}{=}0 \,.
\end{equation}
Note that the rotation matrix from the gauge states to the Higgs mass states also applies to the tadpoles, yielding the relation between the tadpoles $T_i~(i= H, S)$ and $T_{h_i}~(i=1,2)$
\begin{equation}
    \begin{pmatrix}
        T_{h_1} \\T_{h_2}
    \end{pmatrix}
    =
    R(\al)\cdot
    \begin{pmatrix}
        T_{H} \\T_{S}
    \end{pmatrix}\,.
    \label{RENORM::TadpolCondi}
\end{equation}
The one-loop mass counterterm of the Higgs sector is then given by
\begin{equation}
    \mathcal{M}_{h_1 h_2} \rightarrow \mathcal{M}_{h_1 h_2}+\delta
    \mathcal{M}_{h_1 h_2} \;,
\end{equation}
with
\begin{equation}
    \delta \mathcal{M}_{h_1 h_2} =
    \begin{pmatrix}
        \delta m_{h_1}^2 & 0                \\
        0                & \delta m_{h_2}^2
    \end{pmatrix}
    + R(\al)
    \begin{pmatrix}
        \frac{\delta T_{H}}{v} & 0                             \\
        0                           & \frac{\delta T_{S}}{v_S}
    \end{pmatrix}
    R(\al)^T
    \equiv
    \begin{pmatrix}
        \delta m_{h_1}^2 & 0                \\
        0                & \delta m_{h_2}^2
    \end{pmatrix}
    +  \begin{pmatrix}
        \delta T_{h_1 h_1} & \delta T_{h_1 h_2} \\
        \delta T_{h_2 h_1} & \delta T_{h_2 h_2}
    \end{pmatrix}\,.
    \label{RENORM::MASSCOUNTERTERM}
\end{equation}
Equation (\ref{RENORM::MASSCOUNTERTERM}) is strictly expanded to one-loop order, so that terms $\mathcal{O}\cbrak{\delta\alpha \delta T_i}$ are dropped. Applying \OS~ conditions yields $\cbrak{i=1,2}$
\begin{align}
     & \delta m^2_{h_i} = \Re\sbrak{\Sigma_{h_ih_i}(m_{h_i}^2) - \delta T_{h_ih_i}}\,, \\
     & \delta Z_{h_ih_i}  = -
    \Re\sbrak{\frac{\partial\Sigma_{h_ih_i}(p^2)}{\partial
            p^2}}_{p^2=m_{h_i}^2}\,,                                                   \\
     & \delta Z_{h_ih_j} = \frac{2}{m_{h_i}^2-m_{h_j}^2}
    \Re\sbrak{\Sigma_{h_ih_j}(m_{h_j}^2)-\delta T_{h_ih_j}}\,,\quad
    i\neq j\,.\label{RENORM::ZFACTOR}
\end{align}
There is just one DM candidate and therefore the renormalisation constants are defined by
\begin{equation}
    \delta Z_{\X\X} = - \Re\left.\left[\frac{\partial
          \Sigma_{\X\X}(p^2)}{\partial
          p^2}\right]\right|_{p^=\mX^2}\,, \qquad   \delta m^2_{\X} =
    \Re \,  \left[ \Sigma_{\X\X}(p^2 =\mX^2 )
      -\frac{\delta T_S}{v_S}\right] \,,
\end{equation}
with the self-energy $\Sigma_{\X\X}$ of the DM candidate. 

\subsection{Quark Sector}
In the quark sector we assume a diagonal CKM matrix for simplicity. This means we neglect flavor mixing and the OS scheme is applied for each quark individually. The field strength renormalisation constant has to be formulated for the left- and right-handed field of the quarks $\cbrak{q = u,d,s,b,t}$~\cite{Denner:2019vbn}
\begin{eqnarray}
    q_{L/R} \rightarrow \cbrak{1 + \frac{1}{2} \delta Z^{L/R}_{qq}} q \,,
\end{eqnarray}
and the mass counterterm is introduced through
\begin{equation}
    m_q \rightarrow m_q + \delta m_q\,.
\end{equation}
The two-point correlation function of the quarks is written as
\begin{equation}
    \Gamma_{qq}(p) = \ii\cbrak{\slashed{p}-m_q} + \ii \left[\slashed{p}\omega_-\Sigma_{qq}^{L}(p^2)+\slashed{p}\omega_+\Sigma_{qq}^{R}(p^2) + m_q \cbrak{\omega_++\omega_-}\Sigma_{qq}^{S}(p^2)\right]\,,
\end{equation}
where the superscripts $L,R$ and $S$ correspond to the left-,
right-handed and scalar parts of the self-energies, respectively. The
$\omega_{\pm}$ are the left- and right-handed projectors. The full set
of counterterms is then given in terms of the left-/right-handed and
scalar parts of the respective self-energies as
\begin{subequations}    
\begin{eqnarray}
    \delta m_q = \frac{m_q}{2}\Re\left[\Sigma_{qq}^L(m_q^2)+\Sigma_{qq}^R(m_q^2)+2 \Sigma_{qq}^S(m_q^2)\right]\,,\\
    \delta Z_{qq}^{L}  = - \Re\left[\Sigma_{qq}^L(m_q^2)\right] - m_q^2 \Re\left.\left[\frac{\partial \Sigma_{qq}^L(p^2)}{\partial p^2}+\frac{\partial \Sigma_{qq}^R(p^2)}{\partial p^2}+2 \frac{\partial \Sigma_{qq}^S(p^2)}{\partial p^2}\right]\right|_{p^2=m_q^2}\,,\\
    \delta Z_{qq}^{R}  = - \Re\left[\Sigma_{qq}^R(m_q^2)\right] - m_q^2 \Re\left.\left[\frac{\partial \Sigma_{qq}^L(p^2)}{\partial p^2}+\frac{\partial \Sigma_{qq}^R(p^2)}{\partial p^2}+2 \frac{\partial \Sigma_{qq}^S(p^2)}{\partial p^2}\right]\right|_{p^2=m_q^2}\,.
\end{eqnarray}
\end{subequations}

\subsection{Renormalisation of the Mixing Angle}
The rotation Eq.~(\ref{eq:diagM}) of the
interaction states $\Phi_H$ and $\Phi_S$ to the mass eigenstates $h_1$
and $h_2$ introduces the mixing angle $\alpha$ that needs to be
renormalized as well. The renormalisation
of the mixing angles in SM extensions was thoroughly discussed
in~Refs.\cite{Bojarski:2015kra,Krause:2016oke,Denner:2016etu,Krause:2016xku,Krause:2017mal,Altenkamp:2017ldc,Altenkamp:2017kxk,Fox:2017hbw,
    Denner2018,
    Grimus:2018rte,Krause:2018wmo,Krause:2019oar,Krause:2019qwe}. There are many possibilities
  to renormalize the mixing angle. One possibility is to use a
  physical process, like a decay. However, it is known that the usage
  of a process-dependent scheme may yield an unphysically large counterterm~\cite{Krause:2016oke} which in turn leads to extremely large corrections.  In this
work we will use the
KOSY scheme, proposed in~Refs.\cite{Pilaftsis:1997dr, Kanemura:2004mg}, which
connects the angle counterterm with the usual OS counterterms for the
scalar.\footnote{Note, however, that the KOSY scheme can lead
to gauge dependent results~\cite{Krause:2016oke}.}The bare
parameter $\alpha_0$ can be expressed in terms of the
renormalized one, $\alpha$, as
\begin{equation}
    \alpha_0 = \alpha + \delta \alpha\,.
\end{equation}
Considering the field strength renormalisation before the  rotation,
\begin{equation}
    \begin{pmatrix}
        h_1 \\h_2
    \end{pmatrix}
    = R\cbrak{\alpha+\delta\alpha} \sqrt{Z_{\Phi}}\begin{pmatrix}
        \Phi_H \\\Phi_S
    \end{pmatrix}\,,
\end{equation}
and expanding it to strict one-loop order,
\begin{equation}
    R\cbrak{\alpha+\delta\alpha} \sqrt{Z_{\Phi}}
    \begin{pmatrix}
        \Phi_H \\\Phi_S
    \end{pmatrix}
    =\underbrace{R(\delta\alpha) R(\alpha)
        \sqrt{Z_{\Phi}}R(\alpha)^T}_{\overset{!}{=}\sqrt{Z_H}} R(\alpha)
    \begin{pmatrix}
        \Phi_H \\\Phi_S
    \end{pmatrix}
    +\mathcal{O} (\delta\alpha^2)
    = \sqrt{Z_H}
    \begin{pmatrix}
        h_1 \\h_2
    \end{pmatrix}\,,
\end{equation}
yields the field strength renormalisation matrix $\sqrt{Z_H} $
connecting the bare and renormalised fields in the mass basis. Using
the rotation matrix expanded at one-loop order results in
\begin{equation}
    \sqrt{Z_{H}} = R(\delta\alpha)
    \begin{pmatrix}
        1+\frac{\delta Z_{h_1h_1}}{2} & \delta C_{h}                    \\
        \delta C_h                    & 1 + \frac{\delta Z_{h_2h_2}}{2}
    \end{pmatrix}
    \approx
    \begin{pmatrix}
        1+\frac{\delta Z_{h_1h_1}}{2} & \delta C_h +\delta \alpha     \\
        \delta C_h - \delta \alpha    & 1+\frac{\delta Z_{h_2h_2}}{2}
    \end{pmatrix}\,.
\end{equation}
Demanding that the field mixing vanishes on the mass shell is
equivalent to identifying the off-diagonal elements of
$\sqrt{Z_H}$ with those in \cref{renorm::ZfactorHiggs},
\begin{equation}
    \frac{\delta Z_{h_1h_2}}{2} \overset{!}{=} \delta C_h +\delta
    \alpha \qquad\text{and}\qquad \frac{\delta Z_{h_2h_1}}{2}
    \overset{!}{=} \delta C_h -\delta \alpha\,.
\end{equation}
With \cref{RENORM::ZFACTOR} the mixing angle counterterm reads
\begin{eqnarray}
    \delta \alpha & = & \frac{1}{4}\cbrak{\delta Z_{h_1h_2}-\delta Z_{h_2h_1}} \\
    &&=
    \frac{1}{2(m_{h_1}^2-m_{h_2}^2)}\Re\cbrak{\Sigma_{h_1h_2}(m_{h_1}^2)+
        \Sigma_{h_1h_2}(m_{h_2}^2)- 2 \delta T_{h_1h_2}}\,.
\end{eqnarray}
We do not perform a comparison of various
  renormalisation schemes, like a process-dependent, $\MSb$, or the
  KOSY scheme, in this work. We note, however, that in our previous work \cite{Glaus:2019itb}, when comparing these three schemes, we found that only the KOSY scheme led to reasonable NLO 
  predictions.

\subsection{Renormalisation of the Singlet VEV $\vS$ \label{sec:singvevrenorm}}

In the Standard Tadpole (ST) scheme that we are using in this work, there is no need to renormalize the singlet VEV $\vS$. It was shown in Ref.~\cite{Sperling:2013eva} that
when choosing an $R_{\xi}$ gauge in the ST scheme there is no
divergence associated with $\vS$ at one-loop order if
the scalar field obeys a rigid invariance.  In these
SM extensions the singlet field is disconnected from the gauge sector and hence invariant under global gauge transformations. This is exactly the case
for typical extended scalar sectors with a singlet
field, like the complex (or real) singlet extension
of the SM or the Next-to-2-Higgs-Doublet Model, where a
real singlet field is added to the
2-Higgs-Doublet Model. 
We note, however, that in the alternative tadpole scheme as defined in Ref.~\cite{Fleischer:1980ub} for the SM and in Ref.~\cite{Krause:2017mal} for the N2HDM this is no longer true and
a counterterm for $\vS$ is needed. 

\section{Spin-Independent Cross Section in the  \model}
\label{sec:xsection}
In the following, the calculation of the spin-independent
(SI) cross section for the direct detection of
  DM is presented. The starting point is the scattering process of a
DM particle with the nucleon. The effective coupling of this process
is denoted by $\alpha_n$,
\begin{equation}
    \begin{tikzpicture}[baseline={([yshift=-.6ex]v.base)}]
        \begin{feynman}
            \vertex[circle, draw=black, fill=blobColor, minimum size=1.0cm] (v) {\(\alpha_n\)};
            \vertex[left=1.2cm of v] (virtual1);
            \vertex[right=1.2cm of v] (virtual2);
            \vertex[above=0.5cm of virtual1] (chii) {\(\chi\)};
            \vertex[above=0.5cm of virtual2] (chio) {\(\chi\)};
            \vertex[below=0.5cm of virtual1] (qi) {\(n\)};
            \vertex[below=0.5cm of virtual2] (qo) {\(n\)};
            \diagram*{
            (chii) -- [scalar] (v) -- [scalar] (chio),
            (qi) -- (v),
            (v) -- (qo),
            };
        \end{feynman}
    \end{tikzpicture}
    = i\mathcal{A}_n = i\alpha_n \overline{u}_nu_n = i\cdot 2m_n\alpha_n\,,
    \label{eq::effDMnuk}
\end{equation}
where it is additionally  assumed that the momentum of the
nucleon is not altered, that is, the momentum transfer 
between the DM particle and the nucleon is negligible. We can then use the normalisation for the spinors, $ \overline{u}_nu_n = 2m_n$. The DM-nucleon cross section with the interaction in \cref{eq::effDMnuk} is then given by 
\begin{equation}
    \sigma_n = \frac{1}{4\pi} \cbrak{\frac{m_n}{m_n+m_{\X}}}^2 \left|\alpha_n\right|^2\,,
\end{equation}
where $m_n$ corresponds to the nucleon mass and $\mX$ to the DM
mass. Since the nucleon is a bound state the contributions to the
effective DM-nucleon coupling is on the one hand given by the light
valence quarks $\cbrak{q=u,d,s}$ and on the other hand by the gluon
interactions. In order to calculate the cross section the parton basis
is used to describe the interaction between the DM and the
nucleon. The SI DM-nucleon cross section is calculated by taking the related operators
in the non-relativistic limit. 
The parton operator basis forming the most general SI-interactions for
scalar DM is given by \cite{Hisano:2015bma}
\begin{equation}
    \mathcal{L}_{\text{eff}} = \sum_q C_S^q\mathcal{O}^q_S + C^g_S \mathcal{O}^g_S + \sum_q C_T^q \mathcal{O}^q_T\,,
    \label{eq:LeffParton}
\end{equation}
with the operators
\begin{subequations}
    \begin{eqnarray}
        \mathcal{O}_S^q = m_q \X^2 \bar{q}q\,,\\
        \mathcal{O}^g_S = \frac{\alpha_s}{\pi} \X^2 \GG\,,\\
        \mathcal{O}^q_T = \frac{1}{\mX^2} \X^2 \ii \partial^{\mu} \ii \partial^{\nu} \underbrace{\frac{1}{2}\ii\bar{q}\cbrak{\partial_{\mu}\gamma_{\nu}+\partial_{\nu}\gamma_{\nu}-\frac{1}{2}g_{\mu\nu}\slashed{\partial}}q}_{\equiv \mathcal{O}_{\mu\nu}^{q}}\,.
    \end{eqnarray}
    \label{eq::partonoperators}
\end{subequations}
\noindent The operators are built with the DM field $\X$, the quark
spinor $q$ and the gluon field strength tensor
$G_a^{\mu\nu}$ and $\alpha_s$ denotes the
  strong coupling constant.
The operator $\mathcal{O}^q_S$ describes the interaction induced by
the quark-DM interactions and $\mathcal{O}^g_S$ the
  one induced by the gluon-DM interactions. The twist-2 operator $\mathcal{O}_{\mu\nu}^{q}$
also contributes to the SI
cross section due to additional gluon induced interactions. 
Assuming on-shell nucleon states $\ket{n}$, the expectation values of the operators in \cref{eq::partonoperators} can be expressed as \cite{Hisano:2012wm,Young:2009zb,Shifman:1978zn}

\begin{subequations}
    \begin{eqnarray}
        \bra{n}m_q\bar{q}q\ket{n} &\equiv& m_n f^n_q \,,\\
        \bra{n}-\frac{\alpha_s}{12\pi}\GG\ket{n} &\equiv& \frac{2}{27}m_n f_g^n \,,
    \end{eqnarray}
    \label{eq::nucmatrix}
\end{subequations}
with the nucleon matrix elements $f^n_q$ and $f^n_g$ calculated on the lattice. The numerical values for the matrix elements used in the analysis are given in \cref{sec::NumericalSetUp}. \cref{eq::nucmatrix} allows to formulate the effective DM-nucleon coupling $\alpha_n$ in terms of the Wilson coefficients defined in \cref{eq:LeffParton} as \cite{Abe:2018bpo}
\begin{equation}
    \sigma_n = \frac{1}{\pi} \left(\frac{m_n}{\mX+m_n}\right)^2 \left|\sum_{q\,=\,u,\,d,\,s} m_n f_q^n C_S^q - \frac{8}{9} m_n f_g^n C_S^g + \frac{3}{4} m_n \sum_{q\,=\,u,\,d,\,s,\,c,\,b} \left(q^n(2) + \bar{q}^n(2)\right) C_T^q\right|^2\,.
    \label{eq::masterEq}
\end{equation}
The numerical values of the second momenta of the quarks $q^n(2)$ are also given in \cref{sec::NumericalSetUp}. \\
In order to give an estimate of the
DM-nucleon cross section the remaining task is to calculate the Wilson
coefficients $C^{q/g}_{S}$ and  $C^{q}_{T}$ in \cref{eq::masterEq}. 

\subsection{SI Cross Section at Tree Level}
\label{sec::Tree-level}
\begin{figure}[h]
    \centering
    \begin{tikzpicture}[baseline={([yshift=-.6ex]virtual.base)}]
        \definecolor{blobColor}{RGB}{191,191,191}
        \begin{feynman}
            \vertex (v1);
            \vertex[below=0.5cm of v1] (virtual);
            \vertex[below=0.4cm of v1] (blob);
            \vertex[below=0.8cm of virtual] (v2);
            \vertex[above=0.5cm of v2] (triangleTip);
            \vertex[left=0.36 of v2] (triangleLeft);
            \vertex[right=0.36cm of v2] (triangleRight);
            \vertex[left=1.2cm of v1] (chii) {\(\chi\)};
            \vertex[right=1.2cm of v1] (chio) {\(\chi\)};
            \vertex[left=1.2cm of v2] (qi) {\(g\)};
            \vertex[right=1.2cm of v2] (qo) {\(g\)};
            \diagram*{
            (chii) -- [scalar] (v1) -- [scalar] (chio),
            (v1) -- [scalar, edge label=\(h_i\)] (triangleTip),
            (qi) -- [gluon] (triangleLeft),
            (triangleRight) -- [fermion] (triangleLeft),
            (triangleRight) -- [gluon] (qo),
            (triangleLeft) -- [fermion] (triangleTip) -- [fermion] (triangleRight)
            };
        \end{feynman}
    \end{tikzpicture}
    \caption{Interaction of a DM particle and a gluon via a Higgs
      boson mediator and a quark loop.}
    \label{fig::gluonViaQuarkTriangle}
\end{figure}
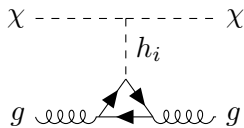

We will start by showing that the SI cross section vanishes in the
limit of vanishing momentum transfer. The Feynman diagrams
representing the quark contributions together with the
corresponding amplitude are given by
\begin{equation}
    \begin{tikzpicture}[baseline={([yshift=-.6ex]virtual.base)}]
        \begin{feynman}
            \vertex (v1);
            \vertex[below=0.5cm of v1] (virtual);
            \vertex[below=1.0cm of v1] (v2);
            \vertex[left=0.7cm of v1] (chii) {\(\chi\)};
            \vertex[right=0.7cm of v1] (chio) {\(\chi\)};
            \vertex[left=0.7cm of v2] (qi) {\(q\)};
            \vertex[right=0.7cm of v2] (qo) {\(q\)};
            \diagram*{
            (chii) -- [scalar] (v1) -- [scalar] (chio),
            (v1) -- [scalar, edge label=\(h_1\)] (v2),
            (qi) -- [fermion] (v2) -- [fermion] (qo),
            };
        \end{feynman}
    \end{tikzpicture}
    +
    \begin{tikzpicture}[baseline={([yshift=-.6ex]virtual.base)}]
        \begin{feynman}
            \vertex (v1);
            \vertex[below=0.5cm of v1] (virtual);
            \vertex[below=1.0cm of v1] (v2);
            \vertex[left=0.7cm of v1] (chii) {\(\chi\)};
            \vertex[right=0.7cm of v1] (chio) {\(\chi\)};
            \vertex[left=0.7cm of v2] (qi) {\(q\)};
            \vertex[right=0.7cm of v2] (qo) {\(q\)};
            \diagram*{
            (chii) -- [scalar] (v1) -- [scalar] (chio),
            (v1) -- [scalar, edge label=\(h_2\)] (v2),
            (qi) -- [fermion] (v2) -- [fermion] (qo),
            };
        \end{feynman}
    \end{tikzpicture}
    = - i \frac{(\mh^2-\mH^2)\cos \alpha \sin \alpha}{(t-\mh^2)(t-\mH^2)v v_S} m_q t \, \bar{u}(p_2) u(p_1)\,, \label{eq::TreeLevelAmplitude}
\end{equation}
where $m_{h_i}~(i=1,2)$ are the neutral Higgs boson masses and
$t=\cbrak{p_{\X}-p_q}$ the Mandelstam variable. The momenta of the DM
particle and the quark are denoted by $p_\chi$ and
  $p_q$, respectively. The amplitude in \cref{eq::TreeLevelAmplitude}
allows to read off the Wilson coefficient $C^q_S$ by identifying $m_q
\bar{u}u$ as the operator $\mathcal{O}^q_S$. The amplitude and
therefore the Wilson coefficient is proportional to the momentum
transfer $t$ and vanishes in the limit of
vanishing momentum transfer. Hence, the
quark contribution to the SI cross section is zero. Note that this
behaviour is related to the $U(1)$ symmetry of the model as will be discussed later. 
Let us show that also the gluon part of the cross section vanishes in
the same limit which implies that the SI cross section vanishes at
tree level in the limit of vanishing momentum
transfer. (Note that the twist-2 operator does not
  contribute at leading order.)\\
The QCD trace anomaly allows to relate the quark operator of the heavy
quarks $Q=b,c,t$ with the gluon field strength tensor yielding the effective gluon interaction with DM particles
\begin{equation}
    m_Q \bar{Q}Q \rightarrow -\frac{\alpha_s}{12 \pi}\GG\,.
    \label{eq::Mapping}
\end{equation}
The corresponding Feynman diagram is depicted in
\cref{fig::gluonViaQuarkTriangle} and can be calculated by first
calculating the process with a (heavy) external quark as in
\cref{eq::TreeLevelAmplitude} and using \cref{eq::Mapping} to
determine the effective gluon interaction. These amplitudes are then used to the extract the Wilson coefficients $C^g_S$. Note that the gluon contributions are extracted in the same way as in \cref{eq::TreeLevelAmplitude}, hence proportional to the momentum transfer $t$. Consequently, also the gluon contributions vanish in the limit of vanishing momentum transfer.
\subsection{EW Corrections to the SI Cross Section}
\label{sec::sec::GenNLO}
As shown in the last section, the SI cross section of the  \model~is suppressed at tree level due to its proportionality to the momentum exchange. Since we work in the limit $t=0$, the tree-level cross section vanishes
and we have to calculate the cross section in the next order of
perturbation theory. In this section we calculate the EW corrections
to the SI cross section which follows very closely our approach
presented in Ref.\cite{Glaus:2019itb} and updated in
Ref.\cite{Glaus:2020ape}. Note that the vector DM model presented in
our previous work does not show the tree-level suppression present in
the \model, hence the cross section is now calculated by taking the NLO amplitude squared
whereas in vector DM model the LO times NLO term was taken.
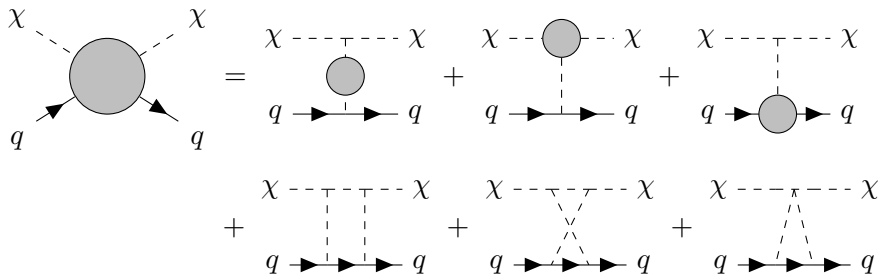
\begin{figure}[t]
    \begin{align*}
        \qquad
        \begin{tikzpicture}[baseline={([yshift=-.6ex]v.base)}]
            \definecolor{blobColor}{RGB}{191,191,191}
            \begin{feynman}
                \vertex[circle, draw=black, fill=blobColor, minimum size=1cm] (v) {};
                \vertex[left=1.2cm of v] (virtual1);
                \vertex[right=1.2cm of v] (virtual2);
                \vertex[above=0.5cm of virtual1] (chii) {\(\chi\)};
                \vertex[above=0.5cm of virtual2] (chio) {\(\chi\)};
                \vertex[below=0.5cm of virtual1] (qi) {\phantom{\(\chi\)}};
                \vertex[below=0.5cm of virtual2] (qo) {\phantom{\(\chi\)}};
                \vertex[below=0.6cm of virtual1] (label1) {\(q\)};
                \vertex[below=0.6cm of virtual2] (label2) {\(q\)};
                \diagram*{
                (chii) -- [scalar] (v) -- [scalar] (chio),
                (qi) -- [fermion] (v),
                (v) --  [fermion] (qo),
                };
            \end{feynman}
        \end{tikzpicture}
         & =
        \begin{tikzpicture}[baseline={([yshift=-.6ex]virtual.base)}]
            \definecolor{blobColor}{RGB}{191,191,191}
            \begin{feynman}
                \vertex (v1);
                \vertex[below=0.5cm of v1] (virtual);
                \vertex[below=0.5cm of virtual] (v2);
                \vertex[left=0.7cm of v1] (chii) {\(\chi\)};
                \vertex[right=0.7cm of v1] (chio) {\(\chi\)};
                \vertex[left=0.7cm of v2] (qi) {\(q\)};
                \vertex[right=0.7cm of v2] (qo) {\(q\)};
                \diagram*{
                (chii) -- [scalar] (v1) -- [scalar] (chio),
                (v1) -- [scalar] (v2),
                (qi) -- [fermion] (v2) -- [fermion] (qo),
                };
            \end{feynman}
            \draw[fill=blobColor](virtual) circle (0.25cm);
        \end{tikzpicture}
        +
        \begin{tikzpicture}[baseline={([yshift=-.6ex]virtual.base)}]
            \definecolor{blobColor}{RGB}{191,191,191}
            \begin{feynman}
                \node[circle, draw=black, fill=blobColor, minimum size=0.5cm] (v1) {};
                \vertex[below=0.5cm of v1] (virtual);
                \vertex[below=1.0cm of v1] (v2);
                \vertex[left=0.95cm of v1] (chii) {\(\chi\)};
                \vertex[right=0.95cm of v1] (chio) {\(\chi\)};
                \vertex[left=0.7cm of v2] (qi) {\(q\)};
                \vertex[right=0.7cm of v2] (qo) {\(q\)};
                \diagram*{
                (chii) -- [scalar] (v1) -- [scalar] (chio),
                (v1) -- [scalar] (v2),
                (qi) -- [fermion] (v2) -- [fermion] (qo),
                };
            \end{feynman}
        \end{tikzpicture}
        +
        \begin{tikzpicture}[baseline={([yshift=-.6ex]virtual.base)}]
            \definecolor{blobColor}{RGB}{191,191,191}
            \begin{feynman}
                \vertex (v1);
                \vertex[below=0.5cm of v1] (virtual);
                \vertex[below=1.0cm of v1] (v2);
                \vertex[left=0.7cm of v1] (chii) {\(\chi\)};
                \vertex[right=0.7cm of v1] (chio) {\(\chi\)};
                \vertex[left=0.7cm of v2] (qi) {\(q\)};
                \vertex[right=0.7cm of v2] (qo) {\(q\)};
                \vertex[left=0.25cm of v2] (qiv);
                \vertex[right=0.25cm of v2] (qov);
                \diagram*{
                (chii) -- [scalar] (v1) -- [scalar] (chio),
                (v1) -- [scalar] (v2),
                (qi) -- [fermion] (qiv) -- (qov) -- [fermion] (qo),
                };
            \end{feynman}
            \draw[fill=blobColor](v2) circle (0.25cm);
        \end{tikzpicture} \nonumber
        \\&+
        \begin{tikzpicture}[baseline={([yshift=-.6ex]virtual.base)}]
            \begin{feynman}
                \vertex (v1);
                \vertex[right=0.5cm of v1] (v1a);
                \vertex[below=1.0cm of v1] (v2);
                \vertex[right=0.5cm of v2] (v2a);
                \vertex[left=0.5cm of v1] (chii) {\(\chi\)};
                \vertex[right=0.5cm of v1a] (chio) {\(\chi\)};
                \vertex[left=0.5cm of v2] (qi) {\(q\)};
                \vertex[right=0.5cm of v2a] (qo) {\(q\)};
                \vertex[below=0.5cm of v1] (virtual);
                \diagram*{
                (chii) -- [scalar] (v1) -- [scalar] (v1a) -- [scalar] (chio),
                (v1) -- [scalar] (v2),
                (v1a) -- [scalar] (v2a),
                (qi) -- [fermion] (v2) -- [fermion] (v2a) -- [fermion] (qo),
                };
            \end{feynman}
        \end{tikzpicture}
        +
        \begin{tikzpicture}[baseline={([yshift=-.6ex]virtual.base)}]
            \begin{feynman}
                \vertex (v1);
                \vertex[right=0.5cm of v1] (v1a);
                \vertex[below=1.0cm of v1] (v2);
                \vertex[right=0.5cm of v2] (v2a);
                \vertex[left=0.5cm of v1] (chii) {\(\chi\)};
                \vertex[right=0.5cm of v1a] (chio) {\(\chi\)};
                \vertex[left=0.5cm of v2] (qi) {\(q\)};
                \vertex[right=0.5cm of v2a] (qo) {\(q\)};
                \vertex[below=0.5cm of v1] (virtual);
                \diagram*{
                (chii) -- [scalar] (v1) -- [scalar] (v1a) -- [scalar] (chio),
                (v1) -- [scalar] (v2a),
                (v1a) -- [scalar] (v2),
                (qi) -- [fermion] (v2) -- [fermion] (v2a) -- [fermion] (qo),
                };
            \end{feynman}
        \end{tikzpicture}
        +
        \begin{tikzpicture}[baseline={([yshift=-.6ex]virtual.base)}]
            \begin{feynman}
                \vertex (v1);
                \vertex[right=0.5cm of v1] (v1a);
                \vertex[right=0.25cm of v1] (v1middle);
                \vertex[below=1.0cm of v1] (v2);
                \vertex[right=0.5cm of v2] (v2a);
                \vertex[left=0.5cm of v1] (chii) {\(\chi\)};
                \vertex[right=0.5cm of v1a] (chio) {\(\chi\)};
                \vertex[left=0.5cm of v2] (qi) {\(q\)};
                \vertex[right=0.5cm of v2a] (qo) {\(q\)};
                \vertex[below=0.5cm of v1] (virtual);
                \diagram*{
                (chii) -- [scalar] (v1) -- [scalar] (v1a) -- [scalar] (chio),
                (v1middle) -- [scalar] (v2),
                (v1middle) -- [scalar] (v2a),
                (qi) -- [fermion] (v2) -- [fermion] (v2a) -- [fermion] (qo),
                };
            \end{feynman}
        \end{tikzpicture}
    \end{align*}
    \caption[One-Loop Electroweak Corrections to DM-Quark Scattering]{One-loop EW corrections to  DM-quark scattering. They are given by propagator corrections, vertex corrections, box and triangle diagrams.}
    \label{fig:allDiagramsQuarks}
\end{figure}
The generic one-loop EW corrections are depicted in
\cref{fig:allDiagramsQuarks}, where the gray blob denotes the
renormalized 4-point vertex (left-side), the renormalized propagator
corrections (first diagram on the right-hand side), the upper
renormalized vertex (second diagram), the lower renormalized vertex
(third diagram) and
the box corrections (last three diagrams). The box
corrections can be split in the genuine square box corrections, crossed box and triangle corrections. In \cref{fig:allDiagramsQuarks} only the quark contributions are shown and we will comment on the gluon contributions at NLO later on.

\subsubsection{Mediator Corrections}
\label{sec::med}
\begin{figure}
    \begin{footnotesize}
        \begin{tikzpicture}[baseline={([yshift=-.6ex]virtual.base)}]
            \begin{feynman}
                \vertex (vmo);
                \vertex[below=0.5cm of vmo] (vm);
                \vertex[below=1.0cm of vmo] (vu);
                \vertex[right=0.55cm of vm] (virtualcircle);
                \vertex[left=0.8cm of vmo] (chii) {\(\X\)};
                \vertex[right=0.8cm of vmo] (chio) {\(\X\)};
                \vertex[left=0.8cm of vu] (qi) {\(q\)};
                \vertex[right=0.8cm of vu] (qo) {\(q\)};
                \vertex[below=0.5cm of vmo] (virtual);
                \vertex[right=0.01cm of virtualcircle] (circlelabel) {\(\Phi_1\)};
                \vertex[left=0.5cm of chii] (vlabel) {a)};
                \diagram*{
                (chii) -- [scalar] (vmo) -- [scalar] (chio),
                (vmo) -- [scalar, edge label'=\(h_i\)] (vm) -- [scalar, edge label'=\(h_j\)] (vu),
                (vm) -- [half right] (virtualcircle),
                (virtualcircle) -- [half right] (vm),
                (qi) -- [fermion] (vu) -- [fermion] (qo),
                };
            \end{feynman}
        \end{tikzpicture}
        \begin{tikzpicture}[baseline={([yshift=-.6ex]virtual.base)}]
            \begin{feynman}
                \vertex (vmo);
                \vertex[below=1.0cm of vmo] (vu);
                \vertex[below=0.3cm of vmo] (vmmo);
                \vertex[above=0.3cm of vu] (vmmu);
                \vertex[left=0.8cm of vmo] (chii) {\(\X\)};
                \vertex[right=0.8cm of vmo] (chio) {\(\X\)};
                \vertex[left=0.8cm of vu] (qi) {\(q\)};
                \vertex[right=0.8cm of vu] (qo) {\(q\)};
                \vertex[below=0.5cm of vmo] (virtual);
                \vertex[left=0.5cm of chii] (vlabel) {b)};
                \diagram*{
                (chii) -- [scalar] (vmo) -- [scalar] (chio),
                (vmo) -- [scalar, edge label'=\(h_i\)] (vmmo),
                (vmmo) -- [scalar, half right, looseness=2.0, edge label'=\(h_k\)] (vmmu),
                (vmmu) -- [scalar, half right, looseness=2.0, edge label'=\(h_l\)] (vmmo),
                (vmmu) -- [scalar, edge label'={\(h_j\)}] (vu),
                (qi) -- [fermion] (vu) -- [fermion] (qo),
                };
            \end{feynman}
        \end{tikzpicture}
        \begin{tikzpicture}[baseline={([yshift=-.6ex]virtual.base)}]
            \begin{feynman}
                \vertex (vmo);
                \vertex[below=1.0cm of vmo] (vu);
                \vertex[below=0.3cm of vmo] (vmmo);
                \vertex[above=0.3cm of vu] (vmmu);
                \vertex[left=0.8cm of vmo] (chii) {\(\X\)};
                \vertex[right=0.8cm of vmo] (chio) {\(\X\)};
                \vertex[left=0.8cm of vu] (qi) {\(q\)};
                \vertex[right=0.8cm of vu] (qo) {\(q\)};
                \vertex[below=0.5cm of vmo] (virtual);
                \vertex[left=0.5cm of chii] (vlabel) {c)};
                \diagram*{
                (chii) -- [scalar] (vmo) -- [scalar] (chio),
                (vmo) -- [scalar, edge label'=\(h_i\)] (vmmo),
                (vmmo) -- [half right, looseness=2.0, edge label'=\(\Phi_2\)] (vmmu),
                (vmmu) -- [half right, looseness=2.0, edge label'=\(\Phi_2\)] (vmmo),
                (vmmu) -- [scalar, edge label'={\(h_j\)}] (vu),
                (qi) -- [fermion] (vu) -- [fermion] (qo),
                };
            \end{feynman}
        \end{tikzpicture}
        \begin{tikzpicture}[baseline={([yshift=-.6ex]virtual.base)}]
            \begin{feynman}
                \vertex (vmo);
                \vertex[below=1.0cm of vmo] (vu);
                \vertex[below=0.3cm of vmo] (vmmo);
                \vertex[above=0.3cm of vu] (vmmu);
                \vertex[left=0.8cm of vmo] (chii) {\(\chi\)};
                \vertex[right=0.8cm of vmo] (chio) {\(\chi\)};
                \vertex[left=0.8cm of vu] (qi) {\(q\)};
                \vertex[right=0.8cm of vu] (qo) {\(q\)};
                \vertex[below=0.5cm of vmo] (virtual);
                \vertex[left=0.5cm of chii] (vlabel) {d)};
                \diagram*{
                (chii) -- [scalar] (vmo) -- [scalar] (chio),
                (vmo) -- [scalar, edge label'=\(h_i\)] (vmmo),
                (vmmo) -- [scalar, half right, looseness=2.0, edge label'={\(G_0\)}] (vmmu),
                (vmmu) -- [boson, half right, looseness=2.0, edge label'={\(Z\)}] (vmmo),
                (vmmu) -- [scalar, edge label'={\(h_j\)}] (vu),
                (qi) -- [fermion] (vu) -- [fermion] (qo),
                };
            \end{feynman}
        \end{tikzpicture}
        \begin{tikzpicture}[baseline={([yshift=-.6ex]virtual.base)}]
            \begin{feynman}
                \vertex (vmo);
                \vertex[below=1.0cm of vmo] (vu);
                \vertex[below=0.3cm of vmo] (vmmo);
                \vertex[above=0.3cm of vu] (vmmu);
                \vertex[left=0.8cm of vmo] (chii) {\(\chi\)};
                \vertex[right=0.8cm of vmo] (chio) {\(\chi\)};
                \vertex[left=0.8cm of vu] (qi) {\(q\)};
                \vertex[right=0.8cm of vu] (qo) {\(q\)};
                \vertex[below=0.5cm of vmo] (virtual);
                \vertex[left=0.5cm of chii] (vlabel) {e)};
                \diagram*{
                (chii) -- [scalar] (vmo) -- [scalar] (chio),
                (vmo) -- [scalar, edge label'=\(h_i\)] (vmmo),
                (vmmo) -- [scalar, half right, looseness=2.0, edge label'={\(G^+\)}] (vmmu),
                (vmmu) -- [boson, half right, looseness=2.0, edge label'={\(W^+\)}] (vmmo),
                (vmmu) -- [scalar, edge label'={\(h_j\)}] (vu),
                (qi) -- [fermion] (vu) -- [fermion] (qo),
                };
            \end{feynman}
        \end{tikzpicture}
        \begin{tikzpicture}[baseline={([yshift=-.6ex]virtual.base)}, cross/.style={path picture={
                                \draw[black]
                                (path picture bounding box.south east) -- (path picture bounding box.north west) (path picture bounding box.south west) -- (path picture bounding box.north east);
                            }}]
            \begin{feynman}
                \vertex (v1);
                \node[below=0.35cm of v1, circle, cross, draw=black, inner sep=0pt, minimum size=0.3cm] (virtual);
                \vertex[below=1.0cm of v1] (v2);
                \vertex[left=0.7cm of v1] (chii) {\(\chi\)};
                \vertex[right=0.7cm of v1] (chio) {\(\chi\)};
                \vertex[left=0.7cm of v2] (qi) {\(q\)};
                \vertex[right=0.7cm of v2] (qo) {\(q\)};
                \vertex[left=0.5cm of chii] (vlabel) {f)};
                \diagram*{
                (chii) -- [scalar] (v1) -- [scalar] (chio),
                (v1) -- [scalar, edge label'=\(h_i\)] (virtual),
                (virtual) -- [scalar, edge label'=\(h_j\)] (v2),
                (qi) -- [fermion] (v2) -- [fermion] (qo),
                };
            \end{feynman}
        \end{tikzpicture}
    \end{footnotesize}
    \caption{The one-loop EW corrections to the mediator. They can be
      split in the genuine one-loop diagrams diagrams
        a)-e)) and the respective 
      counterterm amplitude (diagram f)). The indices
      $i,j,k,l=1,2$ indicates the 
      respective Higgs mediator $h_1,h_2$. The possible field
      insertions are given by
      $\Phi_1= \lbrace h_i,\X,
      G^{(0,\pm)},Z,W^{\pm}\rbrace$ and $\Phi_2=\lbrace \X ,
      G^{(0,\pm)},Z,W^{\pm},\eta_Z,
        \eta_W,f\rbrace$, where $f$ stands for all SM fermions,
     $G^{(0,\pm)}$ for the neutral and charged
        Goldstone bosons, respectively, and $\eta_{Z,W}$ for the ghost fields.}
    \label{fig:PCDiag}
\end{figure}
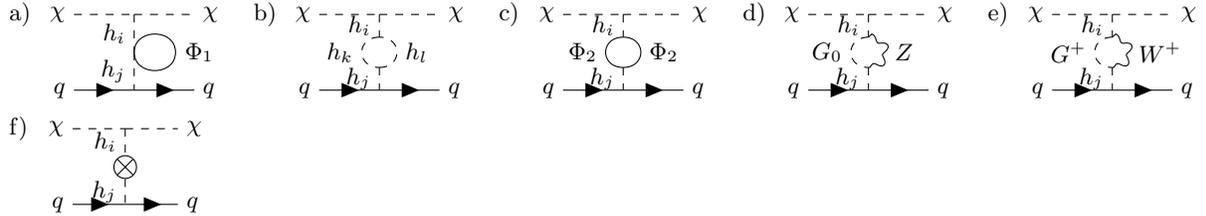

In this section we will discuss the propagator corrections. To
calculate the one-loop corrections to the mediator we first evaluate
all genuine one-loop diagrams in \cref{fig:PCDiag} and construct the
corresponding counterterm. This can be achieved by evaluating the
renormalized one-loop propagator $\cbrak{i,j=1,2}$
\begin{equation}
    \Delta_{h_ih_j} =
    -\frac{\hat{\Sigma}_{h_ih_j}(p^2=0)}{m_{h_i}^2m_{h_j}^2} \;,
\end{equation}
with the renormalised self-energy matrix
\begin{equation}
    \begin{pmatrix}
        \hat{\Sigma}_{h_1h_1} & \hat{\Sigma}_{h_1h_2} \\
        \hat{\Sigma}_{h_2h_1} & \hat{\Sigma}_{h_2h_2}
    \end{pmatrix}
    \equiv
    \hat\Sigma(p^2) = \Sigma(p^2) - \delta m^2 -\delta T +\frac{\delta Z}{2} \cbrak{p^2-\mathcal{M}^2} + \cbrak{p^2-\mathcal{M}^2}\frac{\delta Z}{2}\,.
\end{equation}
We now have everything to determine the contribution of the diagrams
in \cref{fig:PCDiag}. Note that the field strength renormalisation
constant $\delta Z$ is introduced artificially, since the Higgs bosons
correspond to an internal degree of freedom. As it turns out, if the
field strength renormalisation constants are included in all separate
topologies (lower vertex, upper vertex and mediator corrections), they
cancel each other exactly in the sum. Hence, in the end no
artificially introduced $\delta Z$ parts remain in the
calculation. The inclusion of these
$\delta Z$ factors on the other hand
allows to check for the UV finiteness in each topology by itself,
simplifying the calculation or rather the bookkeeping of the contributions. 

\subsubsection{Upper Vertex (upV) Corrections}
\label{sec::upV}
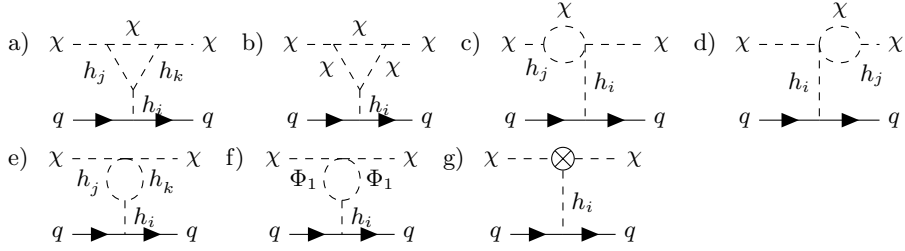
\begin{figure}[]
    \begin{footnotesize}
        \begin{tikzpicture}[baseline={([yshift=-.6ex]virtual.base)}]
            \begin{feynman}
                \vertex (virtualmo);
                \vertex[left=0.35cm of virtualmo] (vlo);
                \vertex[right=0.35cm of virtualmo] (vro);
                \vertex[below=0.6cm of virtualmo] (vm);
                \vertex[below=1.0cm of virtualmo] (vu);
                \vertex[left=0.8cm of virtualmo] (chii) {\(\chi\)};
                \vertex[right=0.8cm of virtualmo] (chio) {\(\chi\)};
                \vertex[left=0.8cm of vu] (qi) {\(q\)};
                \vertex[right=0.8cm of vu] (qo) {\(q\)};
                \vertex[below=0.5cm of virtualmo] (virtual);
                \vertex[left=0.5cm of chii] (vlabel) {a)};
                \diagram*{
                (chii) -- [scalar] (vlo) -- [scalar, edge label=\(\chi\)] (vro) -- [scalar] (chio),
                (vlo) -- [scalar, edge label'=\(h_j\), inner sep=2.0pt, near start] (vm) -- [scalar, edge label'=\(h_k\), inner sep=2.0pt, near end] (vro),
                (vm) -- [scalar, edge label=\(h_i\)] (vu),
                (qi) -- [fermion] (vu) -- [fermion] (qo),
                };
            \end{feynman}
        \end{tikzpicture}
        \begin{tikzpicture}[baseline={([yshift=-.6ex]virtual.base)}]
            \begin{feynman}
                \vertex (virtualmo);
                \vertex[left=0.35cm of virtualmo] (vlo);
                \vertex[right=0.35cm of virtualmo] (vro);
                \vertex[below=0.6cm of virtualmo] (vm);
                \vertex[below=1.0cm of virtualmo] (vu);
                \vertex[left=0.7cm of virtualmo] (chii) {\(\chi\)};
                \vertex[right=0.7cm of virtualmo] (chio) {\(\chi\)};
                \vertex[left=0.7cm of vu] (qi) {\(q\)};
                \vertex[right=0.7cm of vu] (qo) {\(q\)};
                \vertex[below=0.5cm of virtualmo] (virtual);
                \vertex[left=0.5cm of chii] (vlabel) {b)};
                \diagram*{
                (chii) -- [scalar] (vlo) -- [scalar, edge label=\(\chi\)] (vro) -- [scalar] (chio),
                (vlo) -- [scalar, edge label'=\(\chi\),inner sep=2.0pt,near start] (vm) -- [scalar, edge label'=\(\chi\),inner sep=2.0pt, near end] (vro),
                (vm) -- [scalar, edge label=\(h_i\)] (vu),
                (qi) -- [fermion] (vu) -- [fermion] (qo),
                };
            \end{feynman}
        \end{tikzpicture}
        \begin{tikzpicture}[baseline={([yshift=-.6ex]virtual.base)}]
            \begin{feynman}
                \vertex (vmo);
                \vertex[left=0.55cm of vmo] (vlo);
                \vertex[below=1.0cm of vmo] (vu);
                \vertex[left=0.8cm of vmo] (chii) {\(\chi\)};
                \vertex[right=0.8cm of vmo] (chio) {\(\chi\)};
                \vertex[left=0.8cm of vu] (qi) {\(q\)};
                \vertex[right=0.8cm of vu] (qo) {\(q\)};
                \vertex[below=0.5cm of vmo] (virtual);
                \vertex[left=0.5cm of chii] (vlabel) {c)};
                \diagram*{
                (chii) -- [scalar] (vlo) -- [scalar, edge label=\(\chi\), half left] (vmo) -- [scalar] (chio),
                (vlo) -- [scalar, edge label'=\(h_j\), half right, near start, inner sep=0pt] (vmo),
                (vmo) -- [scalar, edge label=\(h_i\)] (vu),
                (qi) -- [fermion] (vu) -- [fermion] (qo),
                };
            \end{feynman}
        \end{tikzpicture}
        \begin{tikzpicture}[baseline={([yshift=-.6ex]virtual.base)}]
            \begin{feynman}
                \vertex (vmo);
                \vertex[right=0.55cm of vmo] (vro);
                \vertex[below=1.0cm of vmo] (vu);
                \vertex[left=0.8cm of vmo] (chii) {\(\chi\)};
                \vertex[right=0.8cm of vmo] (chio) {\(\chi\)};
                \vertex[left=0.8cm of vu] (qi) {\(q\)};
                \vertex[right=0.8cm of vu] (qo) {\(q\)};
                \vertex[below=0.5cm of vmo] (virtual);
                \vertex[left=0.5cm of chii] (vlabel) {d)};
                \diagram*{
                (chii) -- [scalar] (vmo) -- [scalar, edge label=\(\chi\), half left] (vro) -- [scalar] (chio),
                (vmo) -- [scalar, edge label'=\(h_j\), half right, near end, inner sep=2pt] (vro),
                (vmo) -- [scalar, edge label'=\(h_i\)] (vu),
                (qi) -- [fermion] (vu) -- [fermion] (qo),
                };
            \end{feynman}
        \end{tikzpicture}

        \begin{tikzpicture}[baseline={([yshift=-.6ex]virtual.base)}]
            \begin{feynman}
                \vertex (vmo);
                \vertex[below=0.55cm of vmo] (vm);
                \vertex[below=1.0cm of vmo] (vu);
                \vertex[left=0.7cm of vmo] (chii) {\(\chi\)};
                \vertex[right=0.7cm of vmo] (chio) {\(\chi\)};
                \vertex[left=0.7cm of vu] (qi) {\(q\)};
                \vertex[right=0.7cm of vu] (qo) {\(q\)};
                \vertex[below=0.5cm of virtualmo] (virtual);
                \vertex[left=0.5cm of chii] (vlabel) {e)};
                \diagram*{
                (chii) -- [scalar] (vmo) -- [scalar] (chio),
                (vmo) -- [scalar, edge label'=\(h_{j\vphantom{k}}^{\vphantom{Jj}}\),inner sep=2.0pt, half right] (vm),
                (vmo) -- [scalar, edge label=\(h_{k\vphantom{j}}^{\vphantom{Jj}}\),inner sep=2.0pt, half left] (vm),
                (vm) -- [scalar, edge label=\(h_i\)] (vu),
                (qi) -- [fermion] (vu) -- [fermion] (qo),
                };
            \end{feynman}
        \end{tikzpicture}
        \begin{tikzpicture}[baseline={([yshift=-.6ex]virtual.base)}]
            \begin{feynman}
                \vertex (vmo);
                \vertex[below=0.55cm of vmo] (vm);
                \vertex[below=1.0cm of vmo] (vu);
                \vertex[left=0.7cm of vmo] (chii) {\(\chi\)};
                \vertex[right=0.7cm of vmo] (chio) {\(\chi\)};
                \vertex[left=0.7cm of vu] (qi) {\(q\)};
                \vertex[right=0.7cm of vu] (qo) {\(q\)};
                \vertex[below=0.5cm of vmo] (virtual);
                \vertex[left=0.5cm of chii] (vlabel) {f)};
                \diagram*{
                (chii) -- [scalar] (vmo) -- [scalar] (chio),
                (vmo) -- [scalar, edge label'=\( \Phi_{1} \),inner sep=2.0pt, half right] (vm),
                (vm) -- [scalar, edge label'=\( \Phi_{1} \),inner sep=2.0pt, half right] (vmo),
                (vm) -- [scalar, edge label=\(h_i\)] (vu),
                (qi) -- [fermion] (vu) -- [fermion] (qo),
                };
            \end{feynman}
        \end{tikzpicture}
        \begin{tikzpicture}[baseline={([yshift=-.6ex]virtual.base)}, cross/.style={path picture={
                                \draw[black]
                                (path picture bounding box.south east) -- (path picture bounding box.north west) (path picture bounding box.south west) -- (path picture bounding box.north east);
                            }}]
            \begin{feynman}
                \node[circle, cross, draw=black, inner sep=0pt, minimum size=0.3cm] (v1) {};
                \vertex[below=0.5cm of v1] (virtual);
                \vertex[below=1.0cm of v1] (v2);
                \vertex[left=0.95cm of v1] (chii) {\(\chi\)};
                \vertex[right=0.95cm of v1] (chio) {\(\chi\)};
                \vertex[left=0.7cm of v2] (qi) {\(q\)};
                \vertex[right=0.7cm of v2] (qo) {\(q\)};
                \vertex[left=0.5cm of chii] (vlabel) {g)};
                \diagram*{
                (chii) -- [scalar] (v1) -- [scalar] (chio),
                (v1) -- [scalar, edge label=\(h_i\)] (v2),
                (qi) -- [fermion] (v2) -- [fermion] (qo),
                };
            \end{feynman}
        \end{tikzpicture}
        \caption{The EW NLO corrections to the upper vertex. The indices $i,j,k=1,2$ indicate the respective Higgs mediator $h_1,h_2$. The field insertion is given by $\Phi_1=\lbrace \X, G^{(0,\pm)}\rbrace$. }
        \label{fig:VCDiagDM}
    \end{footnotesize}
\end{figure}

The upper vertex corrections - referred to as \upV - are depicted in
\cref{fig:VCDiagDM}. Diagrams a) to f) are the genuine one-loop
corrections and are calculated in the limit of
vanishing momentum transfer 
({\it i.e.}~incoming
momentum is equal to the outgoing momentum, $p_{in}=p_{out}$). Note
that this specific limit is stricter than taking $q^2=\cbrak{p_{\X}-p_q}^2=0$ implying for instance
vanishing Gram determinants complicating the reduction to the standard
one-loop scalar integrals. The numerical evaluation of the
integrals is performed with the {\tt Collier} package
\cite{Denner:2002ii,Denner:2005nn,Denner:2010tr} and explicitly
cross-checked with an in-house 
implementation. The counterterm diagram \cref{fig:VCDiagDM}  g) is
constructed by varying the tree-level coupling of the $\X\X h_i$
vertex $\cbrak{i=1,2}$ 
\begin{equation}
    C_{\X\X h_i} = - \frac{m_{h_i}^2}{v_s} R_{i2}\,,
\end{equation}
yielding
\begin{equation}
    \delta C_{\X\X h_i} = - \frac{R_{i2}}{v_s}\delta
      m_{h_i}^2 - \frac{\delta R_{i2} m_{h_i}^2}{v_s} = - \frac{R_{i2}}{v_s}\delta m_{h_i}^2 - \frac{R_{i1} m_{h_i}^2}{v_s}\delta\alpha\,.
\end{equation}
Note that, since we are using the
standard tadpole scheme, the introduction of a counterterm for the
singlet VEV $v_s$ is not required to obtain a UV finite result
({\it cf.}~discussion in \ref{sec:singvevrenorm}). 
The corresponding counterterm amplitudes read then
\begin{subequations}
    \begin{eqnarray}
        \ii A^{\text{CT}}_{\upV,h_1} = \frac{- C_{qqh_1}}{m_{h_1}^2} \left[\delta C_{\X\X h_1}+\frac{1}{2}\cbrak{C_{\X\X h_1} \delta Z_{h_1h_1}+C_{\X\X h_2}\delta Z_{h_2h_1}}+C_{\X\X h_1}\delta Z_{\X\X}\right]\bar{u}(p_q)u(p_q)\,,\\
        \ii A^{\text{CT}}_{\upV,h_2} = \frac{- C_{qq h_2}}{m_{h_2}^2} \left[\delta C_{\X\X h_2}+\frac{1}{2}\cbrak{C_{\X\X h_2} \delta Z_{h_2h_2}+C_{\X\X h_1}\delta Z_{h_1h_2}}+C_{\X\X h_2}\delta Z_{\X\X}\right]\bar{u}(p_q)u(p_q)\,,
    \end{eqnarray}
\end{subequations}
with the quark Higgs coupling $\cbrak{i=1,2}$
\begin{equation}
    C_{qq h_i } = - \frac{g m_q}{2 m_W} R_{i1}\,,
    \label{eq::lov}
\end{equation}
and the quark spinors $u$. The artificially introduced $\delta Z$ factors for the internal Higgs mediator are again included to ensure the proper cancellation in the sum of all topologies.


\subsubsection{Lower Vertex (loV) Corrections}
\label{sec::lov}
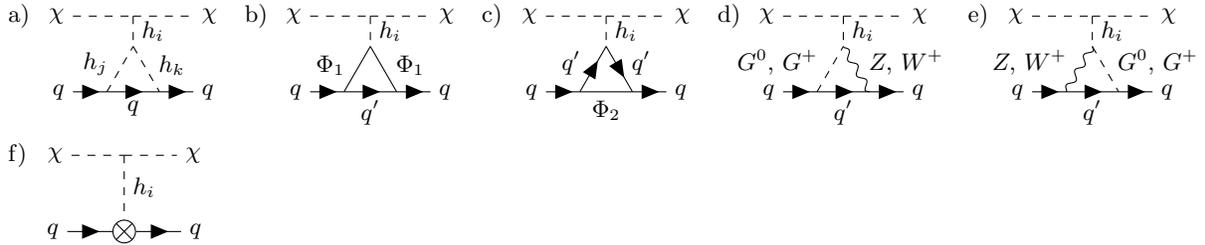
\begin{figure}[]
    \begin{footnotesize}
        \begin{tikzpicture}[baseline={([yshift=-.6ex]virtual.base)}]
            \begin{feynman}
                \vertex (vo);
                \vertex[below=1.0cm of vo] (vu);
                \vertex[left=0.35cm of vu] (vlu);
                \vertex[right=0.35cm of vu] (vru);
                \vertex[below=0.4cm of vo] (vm);
                \vertex[left=0.8cm of vo] (chii) {\(\chi\)};
                \vertex[right=0.8cm of vo] (chio) {\(\chi\)};
                \vertex[left=0.8cm of vu] (qi) {\(q\)};
                \vertex[right=0.8cm of vu] (qo) {\(q\)};
                \vertex[below=0.5cm of vo] (virtual);
                \vertex[left=0.5cm of chii] (vlabel) {a)};
                \diagram*{
                (chii) -- [scalar] (vo) -- [scalar] (chio),
                (vo) -- [scalar, edge label={\(h_i\)}] (vm),
                (vlu) -- [scalar, edge label={\(h_j\)}, inner sep=2.0pt, near start] (vm) -- [scalar, edge label={\(h_k\)}, inner sep=2.0pt, near end] (vru),
                (qi) -- [fermion] (vlu) -- [fermion, edge label'={\(q\)}] (vru) -- [fermion] (qo),
                };
            \end{feynman}
        \end{tikzpicture}
        \begin{tikzpicture}[baseline={([yshift=-.6ex]virtual.base)}]
            \begin{feynman}
                \vertex (vo);
                \vertex[below=1.0cm of vo] (vu);
                \vertex[left=0.35cm of vu] (vlu);
                \vertex[right=0.35cm of vu] (vru);
                \vertex[below=0.4cm of vo] (vm);
                \vertex[left=0.8cm of vo] (chii) {\(\chi\)};
                \vertex[right=0.8cm of vo] (chio) {\(\chi\)};
                \vertex[left=0.8cm of vu] (qi) {\(q\)};
                \vertex[right=0.8cm of vu] (qo) {\(q\)};
                \vertex[below=0.5cm of vo] (virtual);
                \vertex[left=0.5cm of chii] (vlabel) {b)};
                \diagram*{
                (chii) -- [scalar] (vo) -- [scalar] (chio),
                (vo) -- [scalar, edge label={\(h_i\)}] (vm),
                (vlu) -- [edge label={\(\Phi_{1}\)}, inner sep=2.0pt, near start] (vm) -- [edge label={\(\Phi_{1}\)}, inner sep=2.0pt, near end] (vru),
                (qi) -- [fermion] (vlu) -- [fermion, edge label'={\(q'\)}] (vru) -- [fermion] (qo),
                };
            \end{feynman}
        \end{tikzpicture}
        \begin{tikzpicture}[baseline={([yshift=-.6ex]virtual.base)}]
            \begin{feynman}
                \vertex (vo);
                \vertex[below=1.0cm of vo] (vu);
                \vertex[left=0.35cm of vu] (vlu);
                \vertex[right=0.35cm of vu] (vru);
                \vertex[below=0.4cm of vo] (vm);
                \vertex[left=0.8cm of vo] (chii) {\(\chi\)};
                \vertex[right=0.8cm of vo] (chio) {\(\chi\)};
                \vertex[left=0.8cm of vu] (qi) {\(q\)};
                \vertex[right=0.8cm of vu] (qo) {\(q\)};
                \vertex[below=0.5cm of vo] (virtual);
                \vertex[left=0.5cm of chii] (vlabel) {c)};
                \diagram*{
                (chii) -- [scalar] (vo) -- [scalar] (chio),
                (vo) -- [scalar, edge label={\(h_i\)}] (vm),
                (vlu) -- [fermion, edge label={\(q'\)}, inner sep=2.0pt, near start] (vm) -- [fermion, edge label={\(q'\)}, inner sep=2.0pt, near end] (vru),
                (qi) -- [fermion] (vlu) -- [edge label'={\(\Phi_2\)}] (vru) -- [fermion] (qo),
                };
            \end{feynman}
        \end{tikzpicture}
        \begin{tikzpicture}[baseline={([yshift=-.6ex]virtual.base)}]
            \begin{feynman}
                \vertex (vo);
                \vertex[below=1.0cm of vo] (vu);
                \vertex[left=0.35cm of vu] (vlu);
                \vertex[right=0.35cm of vu] (vru);
                \vertex[below=0.4cm of vo] (vm);
                \vertex[left=0.8cm of vo] (chii) {\(\chi\)};
                \vertex[right=0.8cm of vo] (chio) {\(\chi\)};
                \vertex[left=0.8cm of vu] (qi) {\(q\)};
                \vertex[right=0.8cm of vu] (qo) {\(q\)};
                \vertex[below=0.5cm of vo] (virtual);
                \vertex[left=0.5cm of chii] (vlabel) {d)};
                \diagram*{
                (chii) -- [scalar] (vo) -- [scalar] (chio),
                (vo) -- [scalar, edge label={\(h_i\)}] (vm),
                (vlu) -- [scalar, edge label={\(G^0,\,G^+\)}, inner sep=2.0pt, near start] (vm) -- [photon, edge label={\(Z,\,W^+\)}, inner sep=2.0pt, near end] (vru),
                (qi) -- [fermion] (vlu) -- [fermion, edge label'={\(q'\)}] (vru) -- [fermion] (qo),
                };
            \end{feynman}
        \end{tikzpicture}
        \begin{tikzpicture}[baseline={([yshift=-.6ex]virtual.base)}]
            \begin{feynman}
                \vertex (vo);
                \vertex[below=1.0cm of vo] (vu);
                \vertex[left=0.35cm of vu] (vlu);
                \vertex[right=0.35cm of vu] (vru);
                \vertex[below=0.4cm of vo] (vm);
                \vertex[left=0.8cm of vo] (chii) {\(\chi\)};
                \vertex[right=0.8cm of vo] (chio) {\(\chi\)};
                \vertex[left=0.8cm of vu] (qi) {\(q\)};
                \vertex[right=0.8cm of vu] (qo) {\(q\)};
                \vertex[below=0.5cm of vo] (virtual);
                \vertex[left=0.5cm of chii] (vlabel) {e)};
                \diagram*{
                (chii) -- [scalar] (vo) -- [scalar] (chio),
                (vo) -- [scalar, edge label={\(h_i\)}] (vm),
                (vlu) -- [photon, edge label={\(Z,\,W^+\)}, inner sep=2.0pt, near start] (vm) -- [scalar, edge label={\(G^0,\,G^+\)}, inner sep=2.0pt, near end] (vru),
                (qi) -- [fermion] (vlu) -- [fermion, edge label'={\(q'\)}] (vru) -- [fermion] (qo),
                };
            \end{feynman}
        \end{tikzpicture}
        \begin{tikzpicture}[baseline={([yshift=-.6ex]virtual.base)}, cross/.style={path picture={
                                \draw[black]
                                (path picture bounding box.south east) -- (path picture bounding box.north west) (path picture bounding box.south west) -- (path picture bounding box.north east);
                            }}]
            \begin{feynman}
                \vertex (v1);
                \vertex[below=0.5cm of v1] (virtual);
                \node[below=0.85cm of v1, circle, cross, draw=black, inner sep=0pt, minimum size=0.3cm] (v2) {};
                \vertex[left=0.7cm of v1] (chii) {\(\chi\)};
                \vertex[right=0.7cm of v1] (chio) {\(\chi\)};
                \vertex[left=0.95cm of v2] (qi) {\(q\)};
                \vertex[right=0.95cm of v2] (qo) {\(q\)};
                \vertex[left=0.5cm of chii] (vlabel) {f)};
                \diagram*{
                (chii) -- [scalar] (v1) -- [scalar] (chio),
                (v1) -- [scalar, edge label=\(h_i\)] (v2),
                (qi) -- [fermion] (v2) -- [fermion] (qo),
                };
            \end{feynman}
        \end{tikzpicture}
    \end{footnotesize}
    \caption{The EW NLO corrections to the lower vertex. The indices
      $i,j,k=1,2$ indicate the respective Higgs mediator
      $h_1,h_2$. The possible field insertions are given by
      $\Phi_{1}=\lbrace G^{(0,\pm)}, Z , W^{\pm}\rbrace$,
      $\Phi_2=\lbrace h_i,G^{(0,\pm)},\gamma,Z,W^{\pm}\rbrace$. The
      quark $q^{\prime}$ corresponds to the up- or down-type quark
      depending on the field insertion, respectively. Note that for
      simplicity a diagonal CKM matrix is
      assumed.}
    \label{fig:VCDiagQuark}
\end{figure}
\begin{figure}[]
    \centering
    \subfigure{\includegraphics[width=0.25\textwidth]{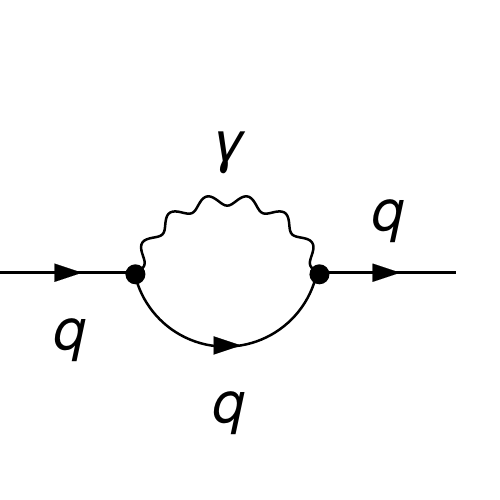}}
    \subfigure{\includegraphics[width=0.25\textwidth]{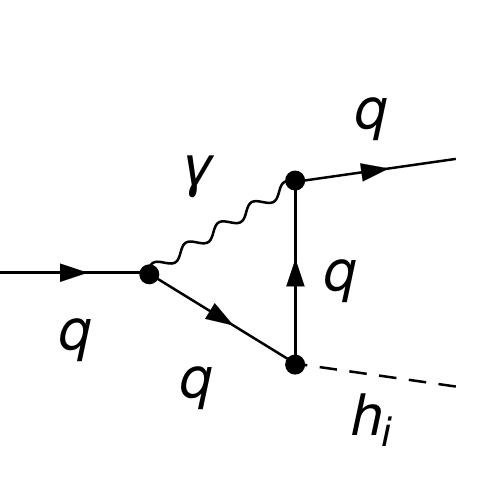}}
    \caption{The QED subset. Left: The quark self-energy containing an internal photon line. Right: The vertex correction with an internal photon line.}
    \label{fig::QEDsubset}
\end{figure}

In \cref{fig:VCDiagQuark} all diagrams needed for the lower vertex, in
the following referred as \loV, are shown. The diagrams (a-e)
correspond to the genuine one-loop diagrams calculated for vanishing
momentum transfer. The counterterm amplitude \cref{fig:VCDiagQuark}
(f) is obtained in the same way as for the \upV. The tree-level
coupling of the lower vertex $qq h_i$ is given in \cref{eq::lov},
hence the counterterm for this vertex reads
\begin{equation}
    \delta C_{qqh_i} = \frac{- gm_q}{2m_W}\cbrak{ R_{i1} \cbrak{-\frac{\delta m_W^2}{2 m_W^2}+\frac{\delta g}{g}+\frac{\delta m_q}{m_q}} - R_{i2} \delta \alpha }\,,
\end{equation}
and the full CT amplitude
\begin{eqnarray}
    \ii\mathcal{A}^{CT}_{\loV,h_1} = \frac{- C_{\X\X h_1}}{\mh^2} \cbrak{\delta C_{qqh_1} +\frac{1}{2}\cbrak{C_{qqh_1}\delta Z_{h_1h_1}+C_{qqh_2}\delta Z_{h_2h_1} + C_{qqh_1}\delta Z^L_{qq}+ C_{qqh_1}\delta Z^R_{qq}} }\,,\\
    \ii\mathcal{A}^{CT}_{\loV,h_2} = \frac{- C_{\X\X h_2}}{\mH^2} \cbrak{\delta C_{qqh_2} +\frac{1}{2}\cbrak{C_{qqh_2}\delta Z_{h_2h_2}+C_{qqh_1}\delta Z_{h_1h_2} + C_{qqh_2}\delta Z^L_{qq}+ C_{qqh_2}\delta Z^R_{qq}} }\,.
\end{eqnarray}
The presence of charged particles in the final states indicates
additional infrared (IR) divergencies in the
amplitudes. The introduction of real radiation to regulate these IR
divergencies does not work in this context, since the matching to the
parton operators in \cref{eq::partonoperators} occurs at the amplitude
level and the cancellation of the IR divergencies happens at the cross
section level. Furthermore, the inclusion of real corrections would
also introduce additional tensor structures in the amplitude which
 have
to be accounted for in the parton operator basis.\\ 
The IR divergent
parts of the amplitude form a closed subset of diagrams referred to
as QED subset in the following and all diagrams contain an internal
photon line. The corresponding diagrams are depicted in
\cref{fig::QEDsubset} where the self-energy of the quarks enters
through the mass counterterm $\delta m_q$ and the field strength
renormalisation constants $\delta Z_{qq}^{L/R}$, and the vertex
corrections are part of the genuine
one-loop vertex corrections of the lower vertex. 
One possible solution is the expansion of the QED
subset in terms of the external quark momentum $p_q$ yielding an IR
safe result as discussed in Ref.\cite{Glaus:2020ape}. However, the
$U(1)$ symmetry of the potential leads to the complete
cancellation of the QED subset, such that no IR divergencies are
present in the final renormalized amplitude of the lower vertex
corrections. Hence no additional treatment is required to regulate IR divergencies. 

\subsubsection{Box Diagrams}
\label{sec::box}
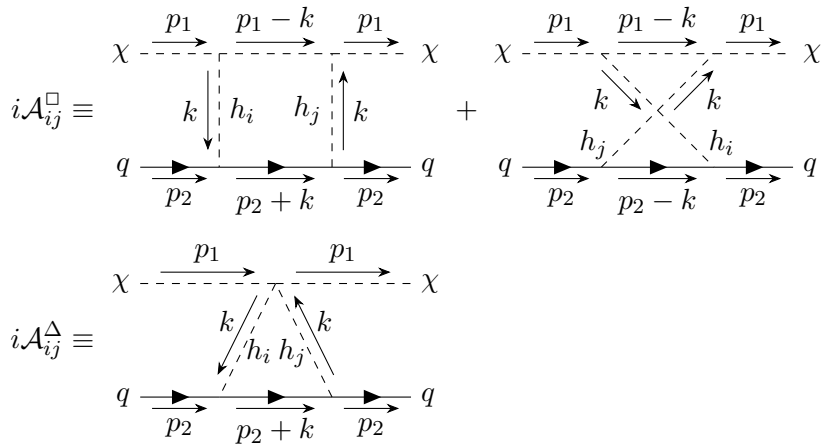
\begin{figure}[h]
    \begin{align*}
        \qquad i \mathcal{A}_{ij}^\Box & \equiv
        \begin{tikzpicture}[baseline={([yshift=-.6ex]virtual.base)}]
            \begin{feynman}
                \vertex (v1);
                \vertex[below=0.75cm of v1] (virtual);
                \vertex[left=0.75cm of v1] (vlo);
                \vertex[right=0.75cm of v1] (vro);
                \vertex[left=1.8cm of v1] (chii) {\(\chi\)};
                \vertex[right=1.8cm of v1] (chio) {\(\chi\)};
                \vertex[below=1.50cm of v1] (v2);
                \vertex[left=1.8cm of v2] (qi) {\(q\)};
                \vertex[right=1.8cm of v2] (qo) {\(q\)};
                \vertex[below=1.50cm of vlo] (vlu);
                \vertex[below=1.50cm of vro] (vru);
                \diagram*{
                (chii) -- [scalar, momentum={[arrow distance=1.5mm]\(p_1\)}] (vlo) -- [scalar, momentum={[arrow distance=1.5mm]\(p_1-k\)}] (vro) -- [scalar, momentum={[arrow distance=1.5mm]\(p_1\)}] (chio),
                (vlo) -- [scalar, momentum'={[arrow distance=1.5mm]\(k\)}, edge label=\(h_i\)] (vlu),
                (vru) -- [scalar, momentum'={[arrow distance=1.5mm]\(k\)}, edge label=\(h_j\)] (vro),
                (qi) -- [fermion, momentum'={[arrow distance=1.5mm]\(p_2\)}] (vlu) -- [fermion, momentum'={[arrow distance=1.5mm]\(p_2+k\)}] (vru) -- [fermion, momentum'={[arrow distance=1.5mm]\(p_2\)}] (qo),
                };
            \end{feynman}
        \end{tikzpicture}
        +
        \begin{tikzpicture}[baseline={([yshift=-.6ex]virtual.base)}]
            \begin{feynman}
                \vertex (v1);
                \vertex[below=0.75cm of v1] (virtual);
                \vertex[left=0.75cm of v1] (vlo);
                \vertex[right=0.75cm of v1] (vro);
                \vertex[left=1.8cm of v1] (chii) {\(\chi\)};
                \vertex[right=1.8cm of v1] (chio) {\(\chi\)};
                \vertex[below=1.50cm of v1] (v2);
                \vertex[left=1.8cm of v2] (qi) {\(q\)};
                \vertex[right=1.8cm of v2] (qo) {\(q\)};
                \vertex[below=1.50cm of vlo] (vlu);
                \vertex[below=1.50cm of vro] (vru);
                \vertex[below=0.75cm of v1] (center);
                \vertex at ($(vlo)+(-0.0cm,-0.6cm)$) (label1) {\(k\)};
                \vertex at ($(vro)+(0.0cm,-0.6cm)$) (label2) {\(k\)};
                \vertex at ($(vlu)+(-0.1cm,0.3cm)$) (label3) {\(h_j\)};
                \vertex at ($(vru)+(0.1cm,0.3cm)$) (label4) {\(h_{i\vphantom{j}}\)};
                \diagram*{
                (chii) -- [scalar, momentum={[arrow distance=1.5mm]\(p_1\)}] (vlo) -- [scalar, momentum={[arrow distance=1.5mm]\(p_1-k\)}] (vro) -- [scalar, momentum={[arrow distance=1.5mm]\(p_1\)}] (chio),
                (vlo) -- [scalar,  momentum'={[arrow distance=1.5mm]}] (center) -- [scalar] (vru),
                (vlu) -- [scalar] (center) -- [scalar,  momentum'={[arrow distance=1.5mm]}] (vro),
                (qi) -- [fermion, momentum'={[arrow distance=1.5mm]\(p_2\)}] (vlu) -- [fermion, momentum'={[arrow distance=1.5mm]\(p_2-k\)}] (vru) -- [fermion, momentum'={[arrow distance=1.5mm]\(p_2\)}] (qo),
                };
            \end{feynman}
        \end{tikzpicture}             \\
        i \mathcal{A}_{ij}^\Delta      & \equiv
        \begin{tikzpicture}[baseline={([yshift=-.6ex]virtual.base)}]
            \begin{feynman}
                \vertex (v1);
                \vertex[below=0.75cm of v1] (virtual);
                \vertex[left=0.75cm of v1] (vlo);
                \vertex[right=0.75cm of v1] (vro);
                \vertex[left=1.8cm of v1] (chii) {\(\chi\)};
                \vertex[right=1.8cm of v1] (chio) {\(\chi\)};
                \vertex[below=1.50cm of v1] (v2);
                \vertex[left=1.8cm of v2] (qi) {\(q\)};
                \vertex[right=1.8cm of v2] (qo) {\(q\)};
                \vertex[below=1.50cm of vlo] (vlu);
                \vertex[below=1.50cm of vro] (vru);
                \vertex at ($(vlu)+(0.1cm,1.0cm)$) (label1) {\(k\)};
                \vertex at ($(vru)+(-0.1cm,1.0cm)$) (label2) {\(k\)};
                \vertex at ($(vlu)+(0.53cm,0.58cm)$) (label3) {\(h_{i\vphantom{j}}\)};
                \vertex at ($(vru)+(-0.53cm,0.58cm)$) (label4) {\(h_j\)};
                \diagram*{
                (chii) -- [scalar, momentum={[arrow distance=1.5mm]\(p_1\)}] (v1) -- [scalar, momentum={[arrow distance=1.5mm]\(p_1\)}] (chio),
                (v1) -- [scalar, momentum'={[arrow distance=1.5mm]}] (vlu),
                (vru) -- [scalar, momentum'={[arrow distance=1.5mm]}] (v1),
                (qi) -- [fermion, momentum'={[arrow distance=1.5mm]\(p_2\)}] (vlu) -- [fermion, momentum'={[arrow distance=1.5mm]\(p_2+k\)}] (vru) -- [fermion, momentum'={[arrow distance=1.5mm]\(p_2\)}] (qo),
                };
            \end{feynman}
        \end{tikzpicture}
    \end{align*}
    \caption{Box topologies contributing to the DM-quark scattering
      referred to as $\mathcal{A}^\Box$ and the
      triangle topologies denoted by
      $\mathcal{A}^\Delta$. The indices $i,j$ denote the Higgs
      mediators $h_1,h_2$.}
    \label{fig::boxTriangleTopologies}
\end{figure}
The box and triangle topologies contributing to the DM-quark
interactions are presented in \cref{fig::boxTriangleTopologies}, where
the incoming momenta are denoted by
$p_1$ and $p_2$, respectively. For simplicity, the triangle diagrams
containing Goldstone bosons ($G^0,G^\pm$) are not shown, but they are
included in the calculation treated in the same way as the Higgs
mediator triangle diagrams in
\cref{fig::boxTriangleTopologies}. 
The definition of the momenta reflecting the vanishing momentum
transfer limit allows to express the diagrams as
\begin{subequations}
    \begin{align}
        \begin{split}
            i \mathcal{A}_{ij}^\Box &= i^4 A_{ij} \, \bar{u}(p_2) \int \frac{d^4k}{(2\pi)^4} \frac{1}{k^2-m_i^2}\frac{1}{(p_1-k)^2-\mX^2}\frac{1}{k^2-m_j^2} \\
            & \hphantom{= i^4 a_i a_j b_i b_j \bar{u}_{p_2} \int \frac{d^4k}{(2\pi)^4}} \cdot \left( \frac{\slashed{p}_2 + \slashed{k} + m_q}{(p_2+k)^2-m_q^2} + \frac{\slashed{p}_2 - \slashed{k} + m_q}{(p_2-k)^2-m_q^2} \right) u(p_2)\,,
        \end{split}                                                                                                                                                                     \\
        i \mathcal{A}_{ij}^\Delta & = i^4 B_{ij} \, \bar{u}(p_2) \int \frac{d^4k}{(2\pi)^4} \frac{1}{k^2-m_i^2}\frac{1}{k^2-m_j^2}\frac{\slashed{p}_2 + \slashed{k} + m_q}{(p_2 + k)^2-m_q^2} u(p_2)\,,
    \end{align}
    \label{eq::BoxTriangleExpr}
\end{subequations}
The generic couplings are defined as $A_{ij}=a_ia_jb_ib_j$ and
$B_{ij}=a_ia_jb_{ij}$, where
$a_{i,j}$ and $b_{i,j}$ are the coefficients of the $h_{i,j}
\bar{q}q$ and $h_{i,j} \chi \chi$ vertices, respectively, and $b_{ij}$ is
the coefficient of the $h_i h_j \chi \chi$ vertex.
The coefficients are given explicitly by
\begin{equation}
    \begin{split}
        & a_1 = -\ii \frac{m_q \cos \alpha}{v}\,, \quad
        a_2 = -\ii \frac{m_q \sin \alpha}{v}\,, \quad
        b_1 = -\ii \frac{\mh^2 \sin\alpha}{\vS}\,, \quad
        b_2 = -\ii \frac{\mH^2 \cos\alpha}{\vS}\,, \\
        & b_{11} = \frac{\sin\alpha}{4v\vS^2} \left(\vS\left(\mH^2-\mh^2\right) \cos^3\alpha + v \mH^2 \cos^2\alpha \sin\alpha + v \mh^2 \sin^3\alpha \right), \\
        & b_{22} = \frac{\cos\alpha}{4v\vS^2} \left(v\mH^2 \cos^3\alpha + v \mh^2 \cos\alpha \sin^2\alpha + \vS \left(\mH^2 - \mh^2\right) \sin^3\alpha \right), \\
        & b_{12} = \frac{\cos\alpha\sin\alpha}{4v\vS^2} \left(2v\mH^2 \cos^2\alpha + 2v \mh^2 \sin^2\alpha - \vS \left(\mH^2 - \mh^2\right) \sin 3\alpha \right).
    \end{split}
    \label{eq::Higgscouplings}
\end{equation}

The main contributions to the amplitudes in \cref{eq::BoxTriangleExpr} come from the regions close to the poles of the propagators, that is where $k^2$ is close to the squared Higgs 
masses $m_1^2$ and $m_2^2$ which are of the order of several
hundreds to thousands of GeV$^2$. In
direct detection experiments, the target nucleus is almost at rest and
hence the energy of the nucleons can be approximated by the Fermi
energy, which is in the order of MeV. Therefore the approximation $p_2
\ll k$ is valid in these integrals and the denominators that contain $p_2$ can be expanded as follows~\cite{Ertas:2019dew, Abe:2018emu},
\begin{equation}
    \frac{1}{(p_2\pm k)^2 - m_q^2} = \frac{1}{k^2 \pm 2p_2\cdot k} = \frac{1}{k^2} \mp \frac{2p_2\cdot k}{k^4} + \mathcal{O}\hspace{-0.2em}\left(\left(\frac{p_2\cdot k}{k^2}\right)^2\right), \label{eq:boxExpansion}
\end{equation}
and using the Dirac equation $\slashed{p} u(p) = m_q u(p)$ we obtain
\begin{subequations}
    \begin{align}
        \ii \mathcal{A}_{ij}^\Box   & = i^4 A_{ij} \, \bar{u}(p_2) \int \frac{d^4k}{(2\pi)^4} \frac{1}{k^2 - m_i^2} \frac{1}{(p_1-k)^2 - \mX^2} \frac{1}{k^2 - m_j^2} \left( \frac{4m_q}{k^2} + \frac{-4p_2\cdot k}{k^4} \slashed{k} \right) u(p_2), \\
        \ii \mathcal{A}_{ij}^\Delta & = i^4 B_{ij} \, \bar{u}(p_2) \int \frac{d^4k}{(2\pi)^4} \frac{1}{k^2-m_i^2} \frac{1}{k^2-m_j^2} \left( \frac{1}{k^2} - \frac{2p_2 \cdot k}{k^4} \right) (2m_q + \slashed{k}) \, u(p_2).
    \end{align}
    \label{eq::PqExpandedAmps}
\end{subequations}
The expanded amplitudes in \cref{eq::PqExpandedAmps} can then be
reduced with standard techniques to the Passarino-Veltmann integral
basis. Furthermore, we emphasise that the expansion leads to reduced scalar
integrals not depending on kinematic variables as $s$ allowing to use
the matching procedure to the parton operator
basis. 

\subsubsection{General Mapping to the Wilson Coefficients}
All diagrams  of the NLO corrections presented in
\cref{sec::med,sec::upV,sec::lov,sec::box} have only two independent
spinor structures contributing to the SI cross section, namely
$\bar{u}(p_2)u(p_2)$ (with the remainder of the amplitude independent
of momenta) and terms proportional to $\cbrak{p_1\cdot p_2}
 \bar{u}(p_2)\slashed{p}_1 u(p_2)$. Hence, the amplitude can be
cast into the following form
\begin{equation}
    \ii \mathcal{A} = \ii \cbrak{A  \bar{u}(p_2)u(p_2) + B\cbrak{p_1\cdot p_2}\bar{u}(p_2)\slashed{p}_1u(p_2)}\,
    \label{eq::EffAmp}
\end{equation}
with some momentum-independent constants $A$ and $B$. The definition of the twist-2 operator allows to reformulate
\begin{equation}
    \bar{q} \ii\partial_{\mu}\gamma_{\nu} = \mathcal{O}^q_{\mu\nu} + \ii \bar{q}\cbrak{\frac{\partial_{\mu}\gamma_\nu-\partial_{\nu}\gamma_\mu}{2}+\frac{1}{4}g_{\mu\nu}\slashed{\partial}}q\,,
\end{equation}
where the asymmetric part does not contribute to the SI cross section and it can therefore be dropped. The resulting amplitude and the coefficients can be mapped to the effective Lagrangian containing the parton operators
\begin{equation}
    \mathcal{L}_{eff} = \cbrak{\frac{1}{2m_q}A+\frac{1}{8}\mX^2B}m_q \X\X\bar{q}q + \frac{1}{2}B \cbrak{\X\ii\partial^{\mu}\ii\partial^\nu\X}\mathcal{O}_{\mu\nu}^q\,.
    \label{eq::EffLagCoeff}
\end{equation}
Identifying the coefficients in \cref{eq::EffLagCoeff} with the Wilson coefficients in \cref{eq:LeffParton} yields
\begin{subequations}
    \begin{eqnarray}
        C^q_S = \frac{1}{2m_q} A + \frac{\mX^2}{8}B\,,\\
        C^q_T = \frac{\mX^2}{2}B\,.
    \end{eqnarray}
    \label{eq::WilsonCoefMap}
\end{subequations}
By using \cref{eq::WilsonCoefMap} the calculated renormalised
amplitude can be mapped to the corresponding Wilson coefficient
allowing to determine the SI cross section at NLO.

\subsubsection{Gluon Contributions}
\label{sec::NLOGluonContributions}
Besides the DM-quark interactions also the DM-gluon interactions contribute to the SI cross section. As shown in \cref{sec::Tree-level} the leading DM-gluon contributions can be obtained by using the relation between the heavy quark operators and the gluon field strength tensor in \cref{eq::Mapping}, but this returns a vanishing SI cross section for vanishing momentum transfer ($t\rightarrow 0$). Therefore, next-to leading order effects have to be taken into account to determine the DM-gluon interactions.
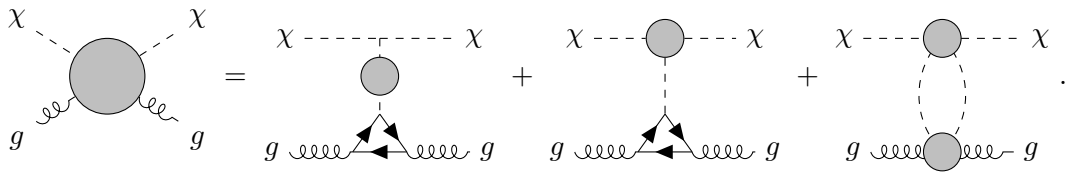
\begin{figure}[h]
    \begin{equation}
        \qquad
        \begin{tikzpicture}[baseline={([yshift=-.6ex]v.base)}]
            \begin{feynman}
                \vertex[circle, draw=black, fill=blobColor, minimum size=1cm] (v) {};
                \vertex[left=1.2cm of v] (virtual1);
                \vertex[right=1.2cm of v] (virtual2);
                \vertex[above=0.5cm of virtual1] (chii) {\(\chi\)};
                \vertex[above=0.5cm of virtual2] (chio) {\(\chi\)};
                \vertex[below=0.5cm of virtual1] (qi) {\phantom{\(\chi\)}};
                \vertex[below=0.5cm of virtual2] (qo) {\phantom{\(\chi\)}};
                \vertex[below=0.6cm of virtual1] (label1) {\(g\)};
                \vertex[below=0.6cm of virtual2] (label2) {\(g\)};
                \diagram*{
                (chii) -- [scalar] (v) -- [scalar] (chio),
                (qi) -- [gluon] (v),
                (v) --  [gluon] (qo),
                };
            \end{feynman}
        \end{tikzpicture}
        =
        \begin{tikzpicture}[baseline={([yshift=-.6ex]virtual.base)}]
            \definecolor{blobColor}{RGB}{191,191,191}
            \begin{feynman}
                \vertex (v1);
                \vertex[below=0.5cm of v1] (virtual);
                \vertex[below=0.5cm of v1] (blob);
                \vertex[below=1cm of virtual] (v2);
                \vertex[above=0.5cm of v2] (triangleTip);
                \vertex[left=0.36cm of v2] (triangleLeft);
                \vertex[right=0.36cm of v2] (triangleRight);
                \vertex[left=1cm of v1] (chii) {\(\chi\)};
                \vertex[right=1cm of v1] (chio) {\(\chi\)};
                \vertex[left=1.2cm of v2] (qi) {\(g\)};
                \vertex[right=1.2cm of v2] (qo) {\(g\)};
                \diagram*{
                (chii) -- [scalar] (v1) -- [scalar] (chio),
                (v1) -- [scalar] (triangleTip),
                (qi) -- [gluon] (triangleLeft),
                (triangleRight) -- [fermion] (triangleLeft),
                (triangleRight) -- [gluon] (qo),
                (triangleLeft) -- [fermion] (triangleTip) -- [fermion] (triangleRight)
                };
            \end{feynman}
            \draw[fill=blobColor](blob) circle (0.25cm);
        \end{tikzpicture}
        +
        \begin{tikzpicture}[baseline={([yshift=-.6ex]virtual.base)}]
            \begin{feynman}
                \node[circle, draw=black, fill=blobColor, minimum size=0.5cm] (v1) {};
                \vertex[below=0.5cm of v1] (virtual);
                \vertex[below=1.5cm of v1] (v2);
                \vertex[above=0.5cm of v2] (triangleTip);
                \vertex[left=0.36cm of v2] (triangleLeft);
                \vertex[right=0.36cm of v2] (triangleRight);
                \vertex[left=1.2cm of v1] (chii) {\(\chi\)};
                \vertex[right=1.2cm of v1] (chio) {\(\chi\)};
                \vertex[left=1.2cm of v2] (qi) {\(g\)};
                \vertex[right=1.2cm of v2] (qo) {\(g\)};
                \diagram*{
                (chii) -- [scalar] (v1) -- [scalar] (chio),
                (v1) -- [scalar] (triangleTip),
                (qi) -- [gluon] (triangleLeft),
                (triangleRight) -- [fermion] (triangleLeft),
                (triangleRight) -- [gluon] (qo),
                (triangleLeft) -- [fermion] (triangleTip) -- [fermion] (triangleRight)
                };
            \end{feynman}
        \end{tikzpicture}
        +
        \begin{tikzpicture}[baseline={([yshift=-.6ex]virtual.base)}]
            \begin{feynman}
                \vertex (v1);
                \vertex[right=0.5cm of v1] (v1a);
                \vertex[right=0.25cm of v1] (v1middle);
                \vertex[below=1.5cm of v1] (v2);
                \vertex[right=0.5cm of v2] (v2a);
                \vertex[right=0.25cm of v2] (v2middle);
                \vertex[left=0.8cm of v1] (chii) {\(\chi\)};
                \vertex[right=0.8cm of v1a] (chio) {\(\chi\)};
                \vertex[left=0.7cm of v2] (qi) {\(g\)};
                \vertex[right=0.7cm of v2a] (qo) {\(g\)};
                \vertex[below=0.5cm of v1] (virtual);
                \diagram*{
                (chii) -- [scalar] (v1) -- [scalar] (v1a) -- [scalar] (chio),
                (v1middle) -- [scalar, quarter left] (v2middle),
                (v1middle) -- [scalar, quarter right] (v2middle),
                (qi) -- [gluon] (v2middle) -- [gluon] (qo),
                };
            \end{feynman}
            \draw[fill=blobColor](v1middle) circle (0.25cm);
            \draw[fill=blobColor](v2middle) circle (0.25cm);
        \end{tikzpicture}. \nonumber
    \end{equation}
    \caption{Generic one-loop correction of the DM-gluon
      interaction. The contributions can be split in mediator, vertex
      corrections and the effective two-loop contributions. The gray
      blob indicates the genuine one-loop corrections and the
      respective counterterm insertion.}
    \label{fig:allDiagramsGluons}
\end{figure}
The leading non-vanishing gluon interactions are shown in
\cref{fig:allDiagramsGluons} which are 2-loop diagrams. The first two
diagrams correspond to the generic mediator and upper vertex EW
corrections in combination with the effective vertex $gg h_i$ which can be calculated in the heavy quark limit (by using \cref{eq::Mapping}). The third term corresponds to an effective two-loop calculation which will be discussed later. \\
The first two diagrams can be calculated using the renormalized upper
vertex (\cref{sec::upV}) and the mediator corrections
(\cref{sec::med}) with external quarks instead of the gluon and using
\cref{eq::Mapping} to effectively determine the DM-gluon
interactions. By identifying the gluon parton operator
$\mathcal{O}^g_s$ the respective Wilson coefficient can be deduced in
accordance with the quark operators. This
method of including the gluon contributions poses several
problems, however.

The first problem is that, as will be shown latter, the correct mass dependence, in the limit $m_\chi \to 0$, is not recovered in the limit of zero DM velocity. 
As discussed in Ref.~\cite{Burgess:1998ku} for an exact symmetry the Goldstone boson completely decouples from all of its interactions in the limit of vanishing momentum. 
Furthermore, it can be shown with the help of a toy model that scattering amplitudes involving Goldstone bosons vanish in the zero-momentum limit
although this is not manifest at the Lagrangian level and only occurs through a nontrivial cancellation of terms in the $S$-matrix. 
The reason is that in the zero-momentum limit the Goldstone state is a symmetry transformation of the ground state and therefore
indistinguishable from the vacuum in this limit~\cite{Burgess:1998ku}. The pseudo Goldstone case is similar - we just have to take simultaneously
the limit of zero-momentum together with $\mX \to 0$ which takes us back to the potential invariant under $U(1)$. 

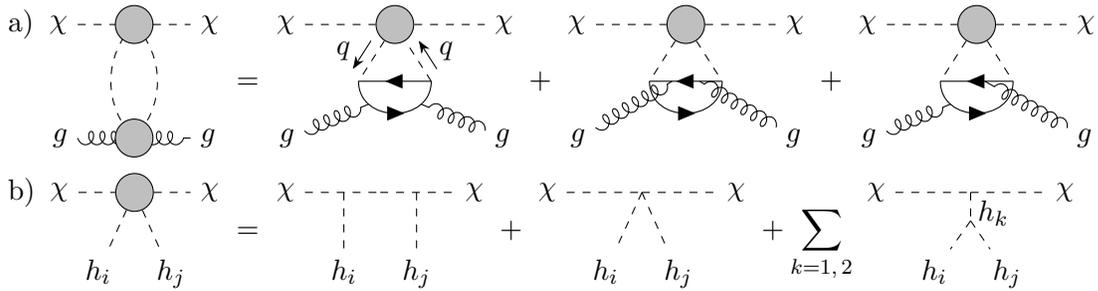
\begin{figure}[h]
    {\setlength{\mathindent}{0.35cm}
        \begin{align*}
             & \begin{tikzpicture}[baseline={([yshift=-.6ex]virtual.base)}]
                \begin{feynman}
                    \vertex (v1);
                    \vertex[right=0.5cm of v1] (v1a);
                    \vertex[right=0.25cm of v1] (v1middle);
                    \vertex[below=1.5cm of v1] (v2);
                    \vertex[right=0.5cm of v2] (v2a);
                    \vertex[right=0.25cm of v2] (v2middle);
                    \vertex[left=0.5cm of v1] (chii) {\(\chi\)};
                    \vertex[right=0.5cm of v1a] (chio) {\(\chi\)};
                    \vertex[left=0.5cm of v2] (qi) {\(g\)};
                    \vertex[right=0.5cm of v2a] (qo) {\(g\)};
                    \vertex[below=0.75cm of v1] (virtual);
                    \vertex[left=0.5cm of chii] (label) {a)};
                    \diagram*{
                    (chii) -- [scalar] (v1) -- [scalar] (v1a) -- [scalar] (chio),
                    (v1middle) -- [scalar, quarter left] (v2middle),
                    (v1middle) -- [scalar, quarter right] (v2middle),
                    (qi) -- [gluon] (v2middle) -- [gluon] (qo),
                    };
                \end{feynman}
                \draw[fill=blobColor](v1middle) circle (0.25cm);
                \draw[fill=blobColor](v2middle) circle (0.25cm);
            \end{tikzpicture}
            =
            \begin{tikzpicture}[baseline={([yshift=-.6ex]virtual.base)}]
                \begin{feynman}
                    \node[circle, draw=black, fill=blobColor, minimum size=0.5cm] (v1);
                    \vertex[below=0.75cm of v1] (virtual);
                    \vertex[left=0.48333cm of v1] (vlo);
                    \vertex[right=0.48333cm of v1] (vro);
                    \vertex[left=1.45cm of v1] (chii) {\(\chi\)};
                    \vertex[right=1.45cm of v1] (chio) {\(\chi\)};
                    \vertex[below=1.5cm of v1] (v2);
                    \vertex[left=1.2cm of v2] (qi) {\(g\)};
                    \vertex[right=1.2cm of v2] (qo) {\(g\)};
                    \vertex[below=0.75cm of vlo] (vlu);
                    \vertex[below=0.75cm of vro] (vru);
                    \vertex at ($(vlu)+(0.14cm,-0.3cm)$) (loop1);
                    \vertex at ($(vru)+(-0.14cm,-0.3cm)$) (loop2);
                    \vertex at ($(vlu)+(-0.2cm,0.4cm)$) (label1) {\(q\)};
                    \vertex at ($(vru)+(0.2cm,0.4cm)$) (label2) {\(q\)};
                    \diagram*{
                    (chii) -- [scalar] (v1) -- [scalar] (chio),
                    (v1) -- [scalar, momentum'={[arrow distance=1.5mm]}] (vlu),
                    (vru) -- [scalar, momentum'={[arrow distance=1.5mm]}] (v1),
                    (vlu) -- [fermion, half right] (vru),
                    (vru) -- [fermion] (vlu),
                    (qi) -- [gluon] (loop1),
                    (qo) -- [gluon] (loop2),
                    };
                \end{feynman}
            \end{tikzpicture}
            +
            \begin{tikzpicture}[baseline={([yshift=-.6ex]virtual.base)}]
                \begin{feynman}
                    \node[circle, draw=black, fill=blobColor, minimum size=0.5cm] (v1);
                    \vertex[below=0.75cm of v1] (virtual);
                    \vertex[left=0.48333cm of v1] (vlo);
                    \vertex[right=0.48333cm of v1] (vro);
                    \vertex[left=1.45cm of v1] (chii) {\(\chi\)};
                    \vertex[right=1.45cm of v1] (chio) {\(\chi\)};
                    \vertex[below=1.5cm of v1] (v2);
                    \vertex[left=1.2cm of v2] (qi) {\(g\)};
                    \vertex[right=1.2cm of v2] (qo) {\(g\)};
                    \vertex[below=0.75cm of vlo] (vlu);
                    \vertex[below=0.75cm of vro] (vru);
                    \vertex[right=0.3cm of vlu] (loop1);
                    \vertex[left=0.3cm of vru] (loop2);
                    \diagram*{
                    (chii) -- [scalar] (v1) -- [scalar] (chio),
                    (v1) -- [scalar] (vlu),
                    (v1) -- [scalar] (vru),
                    (vlu) -- [fermion, half right] (vru),
                    (vru) -- [fermion] (vlu),
                    (qi) -- [gluon] (loop1),
                    (qo) -- [gluon] (loop2),
                    };
                \end{feynman}
            \end{tikzpicture}
            +
            \begin{tikzpicture}[baseline={([yshift=-.6ex]virtual.base)}]
                \begin{feynman}
                    \node[circle, draw=black, fill=blobColor, minimum size=0.5cm] (v1);
                    \vertex[below=0.75cm of v1] (virtual);
                    \vertex[left=0.48333cm of v1] (vlo);
                    \vertex[right=0.48333cm of v1] (vro);
                    \vertex[left=1.45cm of v1] (chii) {\(\chi\)};
                    \vertex[right=1.45cm of v1] (chio) {\(\chi\)};
                    \vertex[below=1.5cm of v1] (v2);
                    \vertex[left=1.2cm of v2] (qi) {\(g\)};
                    \vertex[right=1.2cm of v2] (qo) {\(g\)};
                    \vertex[below=0.75cm of vlo] (vlu);
                    \vertex[below=0.75cm of vro] (vru);
                    \vertex at ($(vlu)+(0.14cm,-0.3cm)$) (loop1);
                    \vertex[left=0.375cm of vru] (loop2);
                    \diagram*{
                    (chii) -- [scalar] (v1) -- [scalar] (chio),
                    (v1) -- [scalar] (vlu),
                    (v1) -- [scalar] (vru),
                    (vlu) -- [fermion, half right] (vru),
                    (vru) -- [fermion] (vlu),
                    (qi) -- [gluon] (loop1),
                    (qo) -- [gluon] (loop2),
                    };
                \end{feynman}
            \end{tikzpicture}    \\
             & \begin{tikzpicture}[baseline={([yshift=-.6ex]virtual.base)}]
                \begin{feynman}
                    \node[circle, draw=black, fill=blobColor, minimum size=0.5cm] (v1);
                    \vertex[below=0.5cm of v1] (virtual);
                    \vertex[left=0.48333cm of v1] (vlo);
                    \vertex[right=0.48333cm of v1] (vro);
                    \vertex[left=1.00cm of v1] (chii) {\(\chi\)};
                    \vertex[right=1.00cm of v1] (chio) {\(\chi\)};
                    \vertex[below=0.75cm of vlo] (vlu) {\(h_i\)};
                    \vertex[below=0.75cm of vro] (vru) {\(h_j\)};
                    \vertex[left=0.5cm of chii] (label) {b)};
                    \diagram*{
                    (chii) -- [scalar] (v1) -- [scalar] (chio),
                    (v1) -- [scalar] (vlu),
                    (v1) -- [scalar] (vru),
                    };
                \end{feynman}
            \end{tikzpicture}
            =
            \begin{tikzpicture}[baseline={([yshift=-.6ex]virtual.base)}]
                \begin{feynman}
                    \vertex (v1);
                    \vertex[below=0.5cm of v1] (virtual);
                    \vertex[left=0.48333cm of v1] (vlo);
                    \vertex[right=0.48333cm of v1] (vro);
                    \vertex[left=1.00cm of v1] (chii) {\(\chi\)};
                    \vertex[right=1.00cm of v1] (chio) {\(\chi\)};
                    \vertex[below=0.75cm of vlo] (vlu) {\(h_i\)};
                    \vertex[below=0.75cm of vro] (vru) {\(h_j\)};
                    \diagram*{
                    (chii) -- [scalar] (v1) -- [scalar] (chio),
                    (vlo) -- [scalar] (vlu),
                    (vro) -- [scalar] (vru),
                    };
                \end{feynman}
            \end{tikzpicture}
            +
            \begin{tikzpicture}[baseline={([yshift=-.6ex]virtual.base)}]
                \begin{feynman}
                    \vertex (v1);
                    \vertex[below=0.5cm of v1] (virtual);
                    \vertex[left=0.48333cm of v1] (vlo);
                    \vertex[right=0.48333cm of v1] (vro);
                    \vertex[left=1.00cm of v1] (chii) {\(\chi\)};
                    \vertex[right=1.00cm of v1] (chio) {\(\chi\)};
                    \vertex[below=0.75cm of vlo] (vlu) {\(h_i\)};
                    \vertex[below=0.75cm of vro] (vru) {\(h_j\)};
                    \diagram*{
                    (chii) -- [scalar] (v1) -- [scalar] (chio),
                    (v1) -- [scalar] (vlu),
                    (v1) -- [scalar] (vru),
                    };
                \end{feynman}
            \end{tikzpicture}
            + \sum_{k=1,\,2}
            \begin{tikzpicture}[baseline={([yshift=-.6ex]virtual.base)}]
                \begin{feynman}
                    \vertex (v1);
                    \vertex[below=0.375cm of v1] (v1u);
                    \vertex[below=0.5cm of v1] (virtual);
                    \vertex[left=0.48333cm of v1] (vlo);
                    \vertex[right=0.48333cm of v1] (vro);
                    \vertex[left=1.00cm of v1] (chii) {\(\chi\)};
                    \vertex[right=1.00cm of v1] (chio) {\(\chi\)};
                    \vertex[below=0.75cm of vlo] (vlu) {\(h_i\)};
                    \vertex[below=0.75cm of vro] (vru) {\(h_j\)};
                    \vertex at ($(v1u)+(0.3cm,+0.1cm)$) (label)  {\(h_k\)};
                    \diagram*{
                    (chii) -- [scalar] (v1) -- [scalar] (chio),
                    (v1) -- [scalar] (v1u),
                    (v1u) -- [scalar] (vlu),
                    (v1u) -- [scalar] (vru),
                    };
                \end{feynman}
            \end{tikzpicture}
        \end{align*}
    } 
    \caption{Triangle and box diagrams with external gluons. a) shows the meaning of the lower blob; b) shows the meaning of the upper blob. }
    \label{fig:gluonBoxQuarkLoop}
\end{figure}

The second problem is that by this matching, only the diagrams with electroweak corrections to the Higgs boson propagator and the upper DM-Higgs boson vertex can be taken into account.
However, electroweak corrections to the lower quark-Higgs boson vertex would obviously interfere with the quark triangle, which makes a matching to heavy quarks non-trivial, since the loops do not factorize. 
This second problem reveals itself in the framework of our renormalisation scheme in the following two points. 
By neglecting the lower vertex corrections, the cancellation of the
artificially introduced $\delta Z_{h_ih_j}$ does not occur anymore and
an uncanceled finite piece of $\delta Z_{h_ih_j}$ remains in the
Wilson coefficient. In particular, the off-diagonal elements of the
wave function renormalisation constants $\delta Z_{h_ih_j}~(i\neq j)$
introduce a mass pole $1/\cbrak{\mH^2-\mh^2}$ yielding a parametric
enhancement for nearly degenerate mass spectra. This divergent
enhancement does not correspond to a physical phenomenon but rather
to a
wrong method for the determination of the DM-gluon interactions. %
Also, the KOSY scheme for the renormalisation of the mixing
counterterm $\delta\alpha$ produces numerically stable (in the sense
of no unphysical parametrically enhanced EW NLO corrections or
divergencies) NLO predictions if either $\delta \alpha$ and $\delta
Z_{h_ih_j}$ appear in a specific combination or if $\delta \alpha$
appears in a full process several times canceling the mass pole
structure\cite{Denner2018}. 
The former occurs \eg in $1\rightarrow 2$
Higgs decays yielding a $\delta Z_{h_ih_j}$ for the on-shell Higgs
state and a corresponding $\delta \alpha$ counterterm in the vertex
counterterm. The latter is present for instance in the $2 \rightarrow
2$ scattering process $\X q \rightarrow \X q$, since $\delta \alpha$
comes both from the upper and from the lower vertices.  
By neglecting in the lower vertex the
  EW corrections in the
 triangle-type diagrams in
\cref{fig:allDiagramsGluons} the conditions for a
numerically stable KOSY mixing counterterm $\delta
\alpha$ are not given, and
hence a non-physical enhancement is expected. 
The third term in \cref{fig:allDiagramsGluons}
 corresponds to an effective two-loop calculation, where the two different gray blobs are explained in \cref{fig:gluonBoxQuarkLoop}.
The diagrams in \cref{fig:gluonBoxQuarkLoop}(a) are calculated using
the approach presented in  Ref.\cite{Ertas:2019dew} and already
applied to the VDM in
Ref.\cite{Glaus:2019itb}. Applying the heavy quark limit (valid for
mediator masses below the top quark mass) allows us to formulate an
effective vertex $h_i h_j gg $ 
\begin{equation}
    \begin{tikzpicture}[baseline={([yshift=-.6ex]v.base)}]
        \begin{feynman}
            \node[circle, draw=black, fill=black, scale=0.5] (v);
            \vertex[left=1.0cm of v] (virtual1);
            \vertex[right=1.0cm of v] (virtual2);
            \vertex[above=0.5cm of virtual1] (chii) {\(h_i\)};
            \vertex[above=0.5cm of virtual2] (chio) {\(h_j\)};
            \vertex[below=0.5cm of virtual1] (qi) {\(g\)};
            \vertex[below=0.5cm of virtual2] (qo) {\(g\)};

            \diagram*{
            (chii) -- [scalar] (v) -- [scalar] (chio),
            (qi) -- [gluon] (v),
            (v) -- [gluon] (qo),
            };
        \end{feynman}
    \end{tikzpicture}
    \approx \frac{ig_s^2}{48\pi^2m_t^2} a_i a_j \,,
\end{equation}
where $a_{i,j}$ are the Higgs-quark couplings defined in \cref{eq::Higgscouplings}. The vertex is produced by the effective Lagrangian
\begin{equation}
    \mathcal{L} \supset \frac{1}{2} \frac{g_s^2}{48\pi^2m_t^2} a_ia_j \, h_ih_jG_{\mu\nu}^a\,G^{a\mu\nu}\,,
\end{equation}
and therefore the Wilson coefficients $\mathcal{O}^g_S$ can be
extracted by calculating the one-loop diagrams induced by the vertices 
depicted in \cref{fig:gluonBoxQuarkLoop}(b). 
The one-loop corrections induced by the last vertex in
\cref{fig:gluonBoxQuarkLoop}(b) have to be calculated with
caution. The first two vertices do not yield a UV pole in the
amplitude, hence no counterterm is required. On the other hand, the
last vertex generates in general a UV pole requiring a vertex
counterterm, since these corrections correspond to an effective vertex
correction. 
  However, the $U(1)$ symmetry of the \model~ensures the cancellation of all UV poles, yielding a UV safe amplitude. 

We emphasise that the inclusion of such effective vertex corrections
has to be done with caution, since the cancellation of the UV poles is
not guaranteed and is model dependent. Furthermore, these corrections
are effective two-loop calculations, where other two-loop
contributions are dropped because they are assumed to be small. This
is not the case in general. Nevertheless, the size of the included
effective two-loop corrections is small compared to the other EW NLO
corrections (\upV,\loV,med,box) when a scan over the allowed parameter
space is performed. Hence, we have included these corrections in our calculation. 
In the following we
will refer to the inclusion of the EW NLO corrections of the upper vertex
or mediator in combination with the effective Higgs-gluon vertex as
the \textit{approach with the additional gluon contributions}. Whereby, the proper SI cross section is calculated
solely by taking the effective two-loop contributions into account
(third diagram of \cref{fig:allDiagramsGluons}. As we will discuss later, these
contributions 
 yield only a sub-percentage effect on the overall cross
section, hence the inclusion of these contributions does not alter the
results significantly.  

\section{Numerical Results}
\label{sec:Res}
\subsection{Numerical Set-Up and Parameters}
\label{sec::NumericalSetUp}
In the following we list the numerical values used for our study. The SM input parameters are taken as \cite{Dittmaier:2011ti}
\begin{equation}
    \begin{split}
        m_u &= \SI{0.19}{GeV}\,,		\qquad &m_c &= \SI{1.4}{GeV}\,,		\qquad &m_t &= \SI{172.5}{GeV}\,, \\
        m_d &= \SI{0.19}{GeV}\,,		\qquad &m_s &= \SI{0.19}{GeV}\,,		\qquad &m_b &= \SI{4.75}{GeV}\,, \\
        m_e &= \SI{0.511}{MeV}\,,		\qquad &m_\mu &= \SI{105.658}{MeV}\,,	\qquad &m_\tau &= \SI{1.777}{GeV}\,, \\
        m_W &= \SI{80.398}{GeV}\,, 	\qquad &v &= \SI{246}{GeV}\,, \\
        m_Z &= \SI{91.188}{GeV}\,. \\
    \end{split}
\end{equation}
The $SU(2)$ electroweak gauge coupling $g$ and the Weinberg angle are expressed in terms of the gauge boson masses and the electroweak VEV
\begin{equation}
    g = 2m_W/v = \SI{0.653}{}\,, \qquad \sin \theta_W = m_W/m_Z = 0.472\,.
\end{equation}
Note that we chose to renormalize the Higgs sector in the mass-ordered Higgs basis $h_1$ and $h_2$ with the masses $\mh<\mH$. One of the Higgs bosons is identified as the SM-like Higgs 
boson with a mass of~\cite{Aad:2015zhl}
\begin{equation}
    m_h = \SI{125.09}{GeV}\,,
\end{equation}
and the non-SM like Higgs boson will be referred to as $\phi$, with mass $m_\phi$. Both mass hierarchies $m_h<m_\phi$ and $m_\phi<m_h$ are allowed in the analysis.

In the following, we refer to the SI cross section as the SI cross section obtained by the scattering on a proton
\begin{equation}
    \sigma  \equiv \sigma_p\,,
\end{equation}
where the proton mass is given by 
\begin{equation}
    m_p = \SI{0.938}{GeV}\,.
\end{equation}
The nuclear matrix elements for the proton needed in \cref{eq::masterEq} are \cite{Hisano:2012wm,Young:2009zb,Shifman:1978zn}  
\begin{equation}
    \begin{split}
        f_u^p &= \SI{0.01513}{}\,,		\qquad &f_d^p &= \SI{0.0191}{}\,,		\qquad &f_s^p &= \SI{0.0447}{}\,, \\
        f_g^p &= \SI{0.92107}{}\,, \\
        u^p(2) &= \SI{0.22}{}\,,		\qquad &c^p(2) &= \SI{0.019}{}\,, \\
        \bar{u}^p(2) &= \SI{0.034}{}\,,		\qquad &\bar{c}^p(2) &= \SI{0.019}{}\,, \\
        d^p(2) &= \SI{0.11}{}\,,		\qquad &s^p(2) &= \SI{0.026}{}\,,	\qquad &b^p(2) &= \SI{0.012}{}\,, \\
        \bar{d}^p(2) &= \SI{0.036}{}\,,		\qquad &\bar{s}^p(2) &= \SI{0.026}{}\,,	\qquad &\bar{b}^p(2) &= \SI{0.012}{}\,,
    \end{split}
\end{equation}
and it should be noted that the uncertainties in the determination of these nuclear matrix elements are not taken into account.
For the parameter region scan we implemented the  \model~in \texttt{ScannerS}~\cite{Coimbra:2013qq, Muhlleitner:2020wwk} which is now publicly available\footnote{The model implementation can be found in \texttt{ScannerS} as CxSMDark.}.

The points generated using \texttt{ScannerS} have to be in agreement with the most relevant experimental and theoretical constraints.  \texttt{ScannerS} allows to check that the potential 
is bounded from below, that there is a global minimum and that perturbative unitarity holds.  The SM-like Higgs couplings to the remaining SM particles are all modified by the same factor. 
Hence, the bound on the signal strength~\cite{Aad:2015zhl} is used to constrain this parameter. 
There are new contributions to the massive gauge-boson self-energies, $\Pi_{WW} (q^2)$ and  $\Pi_{ZZ} (q^2)$. The variables $S, T, U$ \cite{Peskin:1991sw, Grimus:2008nb} 
are used to guarantee agreement with the electroweak precision measurements at the 2$\sigma$ level. 

The collider bounds from LEP, Tevatron and the LHC are all encoded in
\texttt{HiggsBounds 5.6.0}~\cite{Bechtle:2020pkv} and
\texttt{HiggsSignals 2.3.1}~\cite{Bechtle:2013xfa}.  
Agreement at the 95\% confidence level is asked using the exclusion limits for all available searches for non-standard Higgs bosons, including Higgs invisible decays.
The corresponding branching ratios are calculated with
\texttt{AnyHdecay 1.1.0}~\cite{Muhlleitner:2020wwk}. 
This code includes the Higgs decay widths, including the state-of-the
art higher-order QCD corrections, for the complex singlet model
as obtained from~{\tt
  sHDECAY}~\cite{Costa:2015llh}. The code {\tt sHDECAY} is based on
the implementation of the singlet models in
{\tt
  HDECAY}~\cite{Djouadi:1997yw,Djouadi:2018xqq}. For our calculations all EW radiative corrections in
  {\tt HDECAY} are turned off for consistency. 

The DM relic abundance for each model is calculated with the
\texttt{MicrOMEGAs} code~\cite{Belanger:2013oya}, which is compared
with the current experimental result $({\Omega}h^2)^{\rm obs}_{\rm DM}
= 0.1186 \pm 0.002$ from the Planck
Collaboration~\cite{Ade:2015xua}. We do not restrict the DM relic
abundance to be exactly at the experimental value but rather that the value predicted
  by the model has to be equal to or smaller than the observed
central value plus 2$\sigma$. This way, we can consider both the dominant and
subdominant DM cases simultaneously. 
Regarding direct detection the
XENON1T~\cite{Aprile:2017iyp,Aprile:2018dbl} experiment gives the most
stringent upper bound for the DM nucleon scattering.

The scan ranges are chosen to be 
\begin{equation}
    \begin{split}
        m_\phi &\in [&\SI{30}{GeV},&&\SI{1000}{GeV}\,]\,,\\
        m_\chi &\in [&\SI{30}{GeV},&&\SI{1000}{GeV}\,]\,,\\
        v_S &\in [&\SI{1}{GeV},&&\SI{1000}{GeV}\,]\,,\\
        \alpha &\in [&-\pi/2,&&\pi/2\,]\,.
    \end{split} \label{eq:ScannerSRanges}
\end{equation}

\subsection{Results and Discussion}

\begin{figure}[h]
    \centering
    \includegraphics[width=0.8\textwidth]{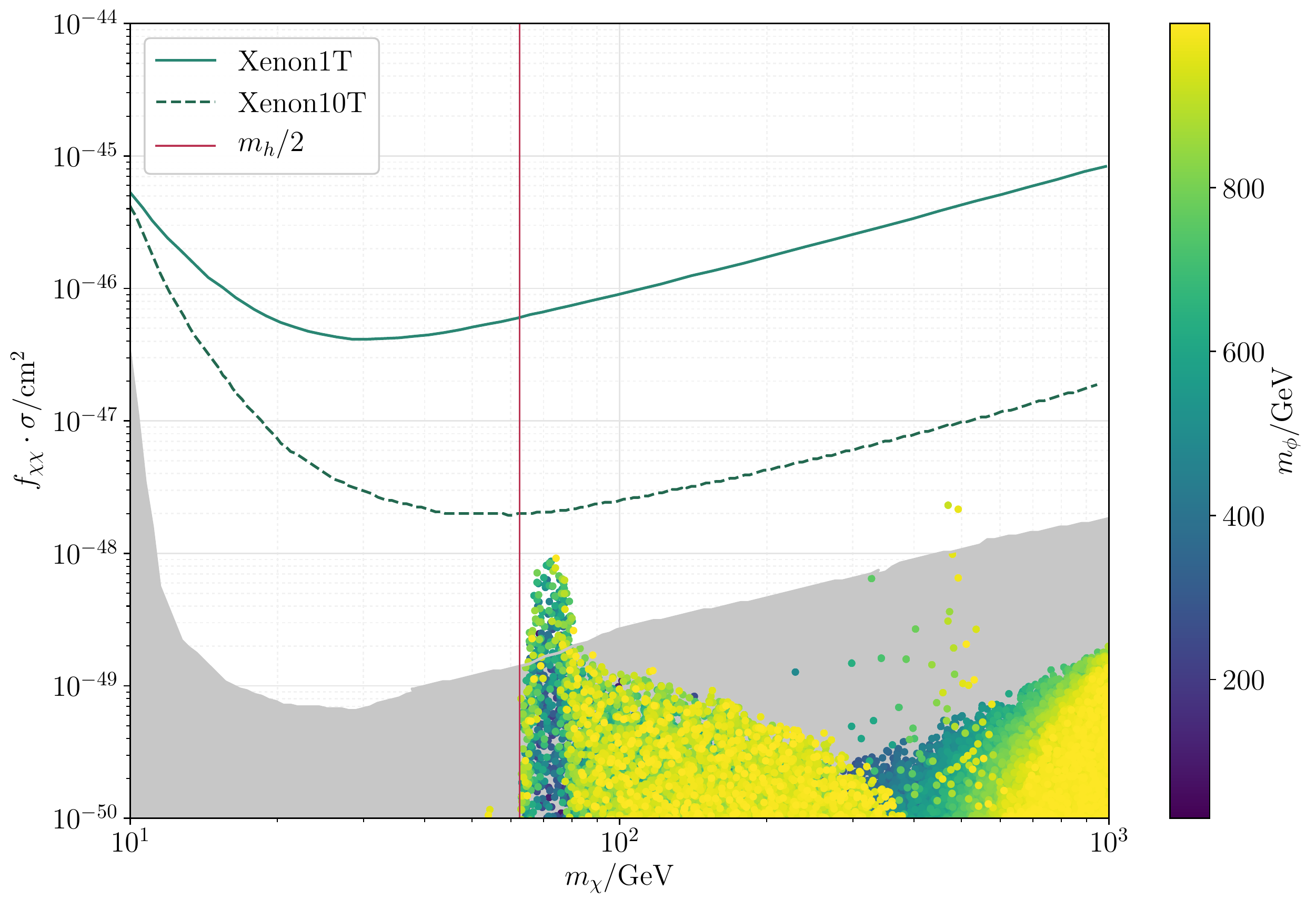}
    \caption{The effective SI-DM nucleon cross section versus the DM
      mass $\mX$ is shown, where the color code indicates
      the value of the non-SM like Higgs boson mass
      $m_{\phi}$. The gray shaded region denotes the neutrino floor
      background and the lines the respective (expected) limits of the
      different experiments. The vertical red line corresponds to $\mX
      = m_h/2$.}
    \label{fig::XenonPlot}
\end{figure}
We start the discussion with the Xenon plot in \cref{fig::XenonPlot}. 
The effective SI DM-nucleon cross section is shown as a function of the DM mass $\mX$. Note that the actual SI cross section has to be rescaled with the factor 
\begin{equation}
    f_{\X\X} \equiv \frac{\cbrak{\Omega h^2}_{\X}}{\cbrak{\Omega h^2}_{\text{DM}}}\,
    \label{eq::CorrectionFactor}
\end{equation}
with the observed relic density $\cbrak{\Omega h^2}_{\text{DM}}$ and
the produced relic density $\cbrak{\Omega h^2}_{\X}$ for the DM WIMP
$\X$. As discussed, we do not demand that the DM candidate accounts
for the full relic density. When DM is under abundant, the effective
factor in \cref{eq::CorrectionFactor} corrects the cross section
accordingly. The relic density is calculated in the standard
freeze-out mechanism with the help of {\tt MicrOmegas} implemented in
{\tt ScannerS}. The color code in \cref{fig::XenonPlot} denotes
the value of the non-SM like Higgs boson mass
$m_\phi$ and the gray shaded region corresponds to the neutrino floor~\cite{Billard:2013qya}. The different lines correspond to the limits of the different DM detection experiments.
The vertical red line indicates the half of the SM-like Higgs boson mass.

All parameter points shown in \cref{fig::XenonPlot} are compatible
with the theoretical and experimental constraints described previously. 
The figure shows that for the entire range of the DM
mass from roughly $40\gev$ up to $1\tev$, only small
mass regions around $\mX \approx m_h/2$ and $\mX \approx m_\phi/2$
may yield an effective SI cross section above the neutrino floor. 
In the case of $\mX \approx m_h/2$ we can see a large number of points that
are basically above but close $m_h/2$; points below  $m_h/2$ are excluded
by the LHC invisible decays constraints. For the region where $\mX \approx m_\phi/2$
only a few points for $m_\phi$ of the order of $1\tev$ are above the neutrino floor.
There are however many points in this region that are above the region where
most points are concentrated. The fact that only scattered points appear in this region
is related to a combination of the experimental constraints.
These regions correspond to the two resonances $h$ and $\phi$, respectively. 
The requirement of proper dark matter abundance leads to the suppression of the coupling between DM and the
resonance. However, the kinematical enhancement caused by the resonance 
compensates for the suppressed couplings that govern DM annihilation in the early Universe. 
Parameter points below the neutrino floor are not of interest, since
those points wiil not be able to be checked by 
future direct detection experiments, as the neutrino floor puts a natural limit to the sensitivity of this kind of experiments. 
The abrupt cut for $\mX$ below $m_h/2$ is induced by Higgs to
invisible searches yielding a strict limit, since in this parameter
region the decay $h\rightarrow \X\X$ is kinematically allowed. Hence,
only a few allowed points are found in this specific parameter region.
We emphasise that the tree-level prediction for the SI cross section
is zero due to the vanishing momentum-transfer limit, hence the
parameter points cannot be constrained by direct detection experiments
with tree-level calculations. However, as shown in
\cref{fig::XenonPlot} the EW NLO corrections can
shift the parameter points above the neutrino floor and in the reach
of the future Xenon 100T experiment. Therefore, the EW NLO corrections might 
play an important role in the discussion of the sensitivity of the direct
detection experiments and derived exclusion limits. \\ 

\begin{figure}[h]
    \centering
    \includegraphics[width=0.49\textwidth]{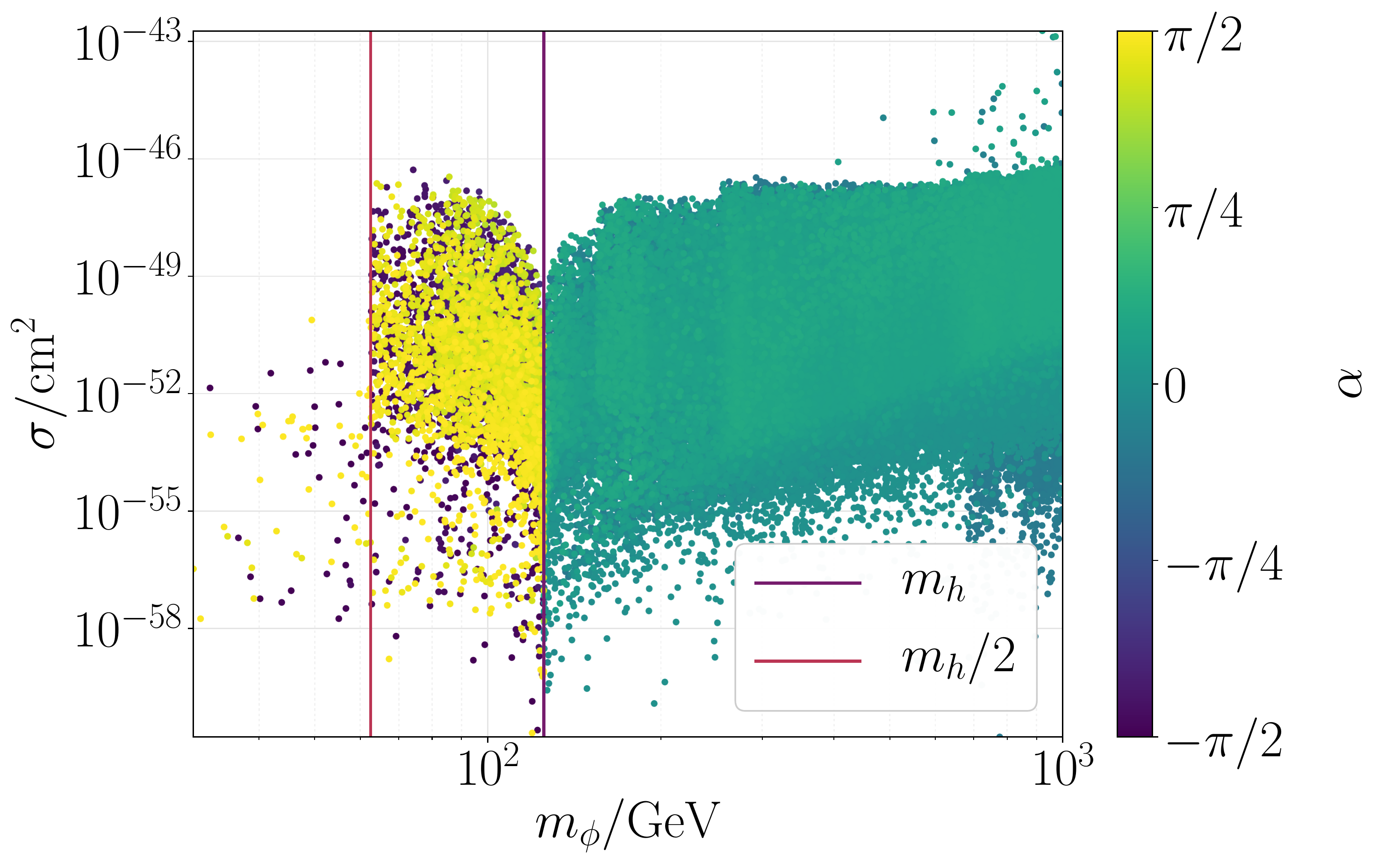}
    \includegraphics[width=0.49\textwidth]{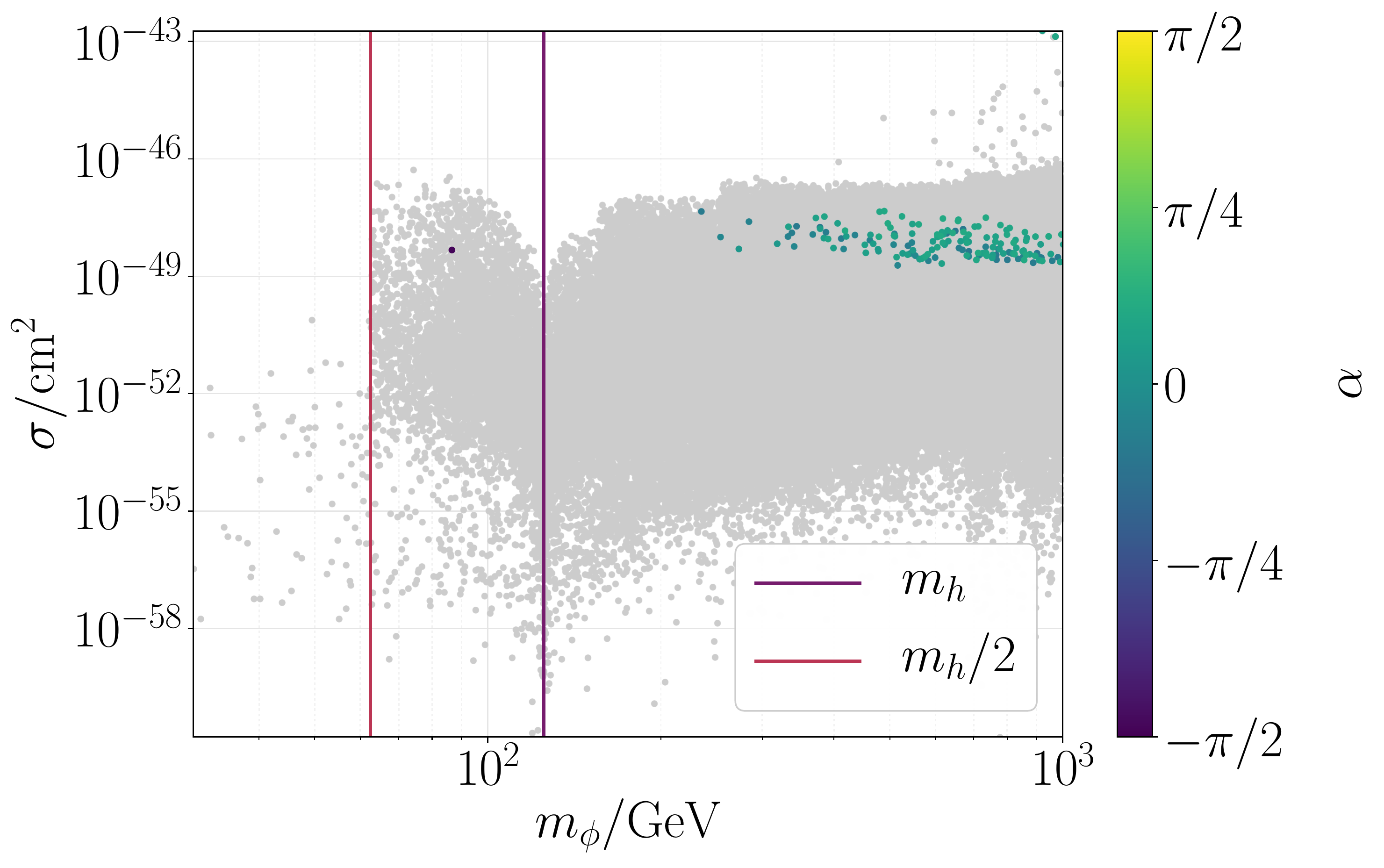}    
    \caption{The SI cross section is shown as a function of the non-SM
      like Higgs boson mass $m_\phi$. The gray points denote the full
      sample passing all experimental and theoretical constraints. The
      colored points yield SI cross sections 
      above the
    neutrino floor, where the color code indicates the mixing angle
    $\alpha$. The two vertical lines indicate $m_\phi=m_h/2$ and
    $m_\phi=m_h$, respectively. On the left-hand side, all of the
    about 260.000 parameter points fulfilling the theoretical and
    experimental constraints are plotted. On the right-hand side, only
    the parameter points that appear above the neutrino floor are
    plotted in color and all remaining parameter points are shown in
    gray.} 
    \label{fig::SIcxnMPHI}
\end{figure}

In \cref{fig::SIcxnMPHI} the SI cross section is shown as a function
of the non-SM like Higgs boson mass $m_\phi$ with the color code
indicating the mixing angle $\alpha$. Note that we do
  not include the factor $f_{\chi\chi}$ here. The SI cross section drops for
degenerate neutral Higgs boson masses ($m_\phi\approx m_h$) because
the NLO cross section is proportional to $m_\phi^2 -m_h^2$ as shown in
Ref.~\cite{Azevedo:2018exj}. 
On the left-hand side, all of the about 260.000 parameter points
fulfilling the theoretical and experimental constraints are
plotted. On the right-hand side, only the parameter points that
lead to direct detection cross sections above the neutrino floor are plotted
in color and all remaining parameter points are shown in gray. It is
interesting to note that there are allowed points with very large
cross sections which, however, do not fulfil the relic density
constraints. This way most points with the appropriate 
relic density have a cross section below $\sim 10^{-46}\text{cm}^2$,
except for a few very heavy non-SM like Higgs boson masses.  
All parameter points with an SI cross section above the neutrino floor
have a maximal mixing between the Higgs doublet gauge state $\Phi_H$
and the singlet $\Phi_S$. Only a single parameter point is above the
neutrino floor with one neutral Higgs boson being a
\textit{singlet-like} Higgs boson, meaning that the mass eigenstate is
almost given by the singlet field component.
 This parameter point is
also the only parameter point having an inverted Higgs spectrum
($m_\phi< m_h$) while providing an SI cross section above the neutrino
floor.

\begin{figure}[h]
    \centering
    \subfigure{\includegraphics[width=0.49\textwidth]{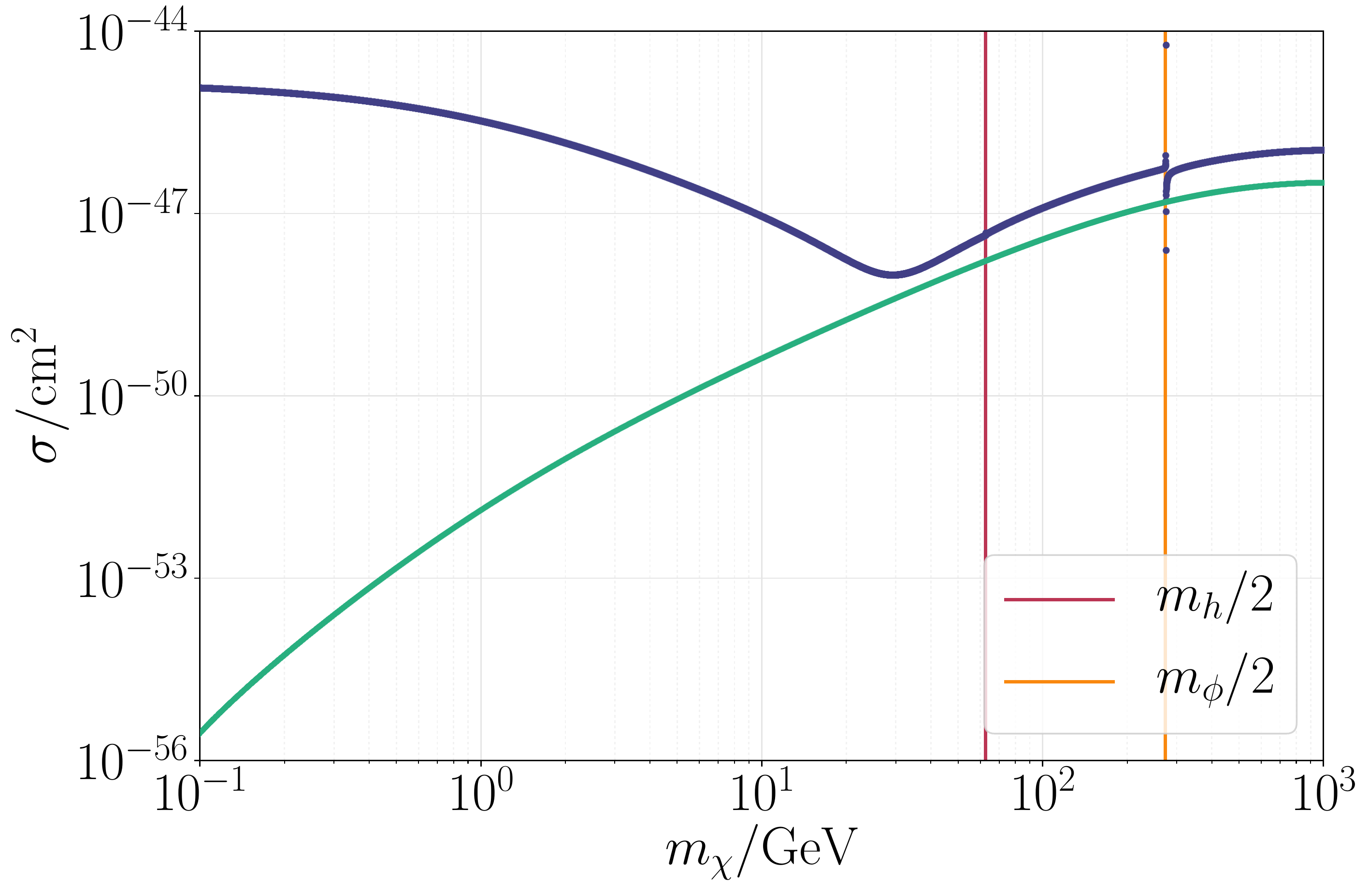}}
    \subfigure{\includegraphics[width=0.49\textwidth]{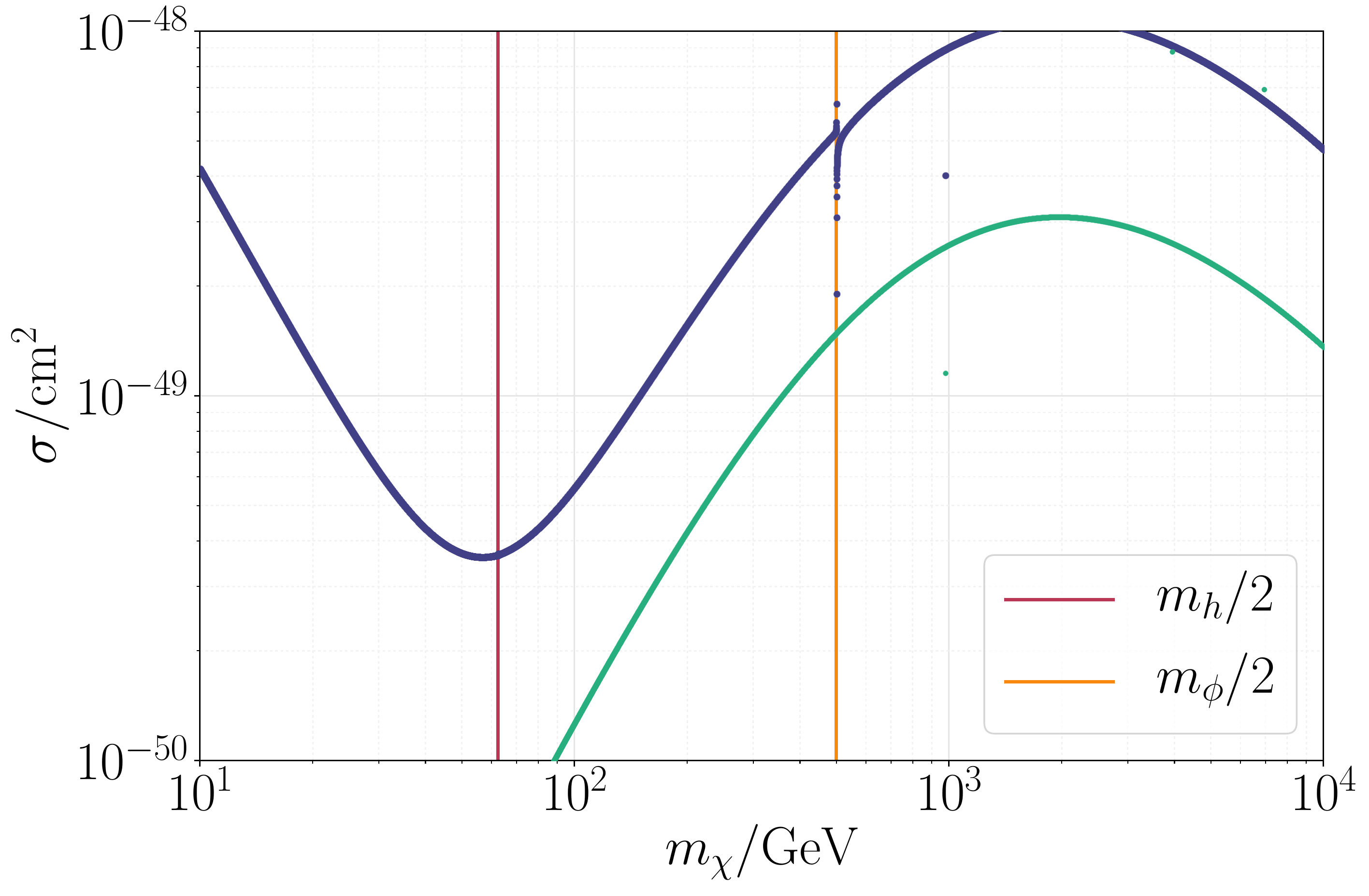}}
    \caption{In both figures the SI cross section is shown as a
      function of the DM mass $\mX$. The green line indicates the
      result calculated in our presented approach and the blue line
      corresponds to the approach including the additional gluon
      contributions discussed in \cref{sec::NLOGluonContributions}. In
      the left plot, our starting point is the
        benchmark given in Tab.~\ref{tab::BMP} and on the right side
      the result using as starting from the benchmark
      point presented in \protect\cite{Ishiwata:2018sdi} with
      $m_\phi=1\tev, \vS = 2v$ and $\sin\alpha=0.2$ is
        shown. 
    }
    \label{fig::CompApproaches}
\end{figure}
In \cref{fig::CompApproaches} we show the SI cross section as a
function of the DM mass $\mX$, where the DM mass is varied while
keeping the other input parameters fixed. On the left side the
resulting SI cross section is shown by starting from
  the benchmark point given in \cref{tab::BMP} and then varying only
  the DM mass while keeping all other parameters fixed, and on the
right side we show the results by starting from the benchmark scenario presented in Ref.\cite{Ishiwata:2018sdi} with the input parameters
\begin{eqnarray}
    m_\phi = 1\tev\,,\quad & \vS = 2 v \,, \quad & \sin\alpha = 0.2
\end{eqnarray}
and variable DM mass.

\begin{table}[h]
    \centering
    \begin{tabular}{c c c c c c}
        \toprule
        $m\phi~\left[\gev\right]$ & $\mX~\left[\gev\right]$ & $\vS~\left[\gev\right]$ & $\alpha$ &$f_{\X\X}$ & $\sigma^{(\text{SI})}~\left[\text{cm}^2\right]$\\\midrule
        546.93 & 72.53 & 152.05 & 0.224 & 0.40 & $8.63\cdot 10^{-49}$\\\bottomrule
    \end{tabular}
    \caption{Benchmark point of the \model: The benchmark point is used to illustrate the parameter dependencies in the following. This parameter point provides an SI cross section above the neutrino floor.}
    \label{tab::BMP}
\end{table}

The green line corresponds to the SI cross section calculated in the
approach presented in \cref{sec::sec::GenNLO} and the blue line shows
the result for the approach with the additional inclusion of the
gluon contributions presented in \cref{sec::NLOGluonContributions}. As
discussed  in \cref{sec::NLOGluonContributions} the additional gluon
contributions induce several problems. The first problem can be clearly seen in 
both plots in  \cref{fig::CompApproaches}. The Goldstone nature of the DM candidate $\X$ requires that the SI cross section scales with the corresponding DM mass $\mX$ \cite{Azevedo:2018exj}, 
implying that the SI cross section vanishes in the zero DM mass limit, since the Goldstone nature of the DM candidate is restored. Note that this particular behaviour is only expected for vanishing momentum transfer as assumed in the calculation. Our approach (neglecting the gluon contributions) shows for both benchmark points (left and right in \cref{fig::CompApproaches}) the desired behaviour for small DM masses $\mX$ which does not happen when the additional gluon contributions to the SI cross section are included.

As for the second problem related to the approximation performed in the
two-loop diagrams, it is not clear how it would reflect on the results. What we can see from the plots 
is that for large DM masses both approaches yield similar results. The difference is roughly a factor three induced by the inclusion of the gluon contributions. Further, the results presented in Ref.\cite{Ishiwata:2018sdi} are exactly reproduced only if we include the additional gluon contributions. The important point here is to understand that unless a complete 2-loop calculation of the gluon contribution 
is performed, nothing can be said about the inclusion of approximate
calculations of some diagrams. 

%

\begin{figure}[h]
    \centering
    \subfigure{\includegraphics[width=0.49\textwidth]{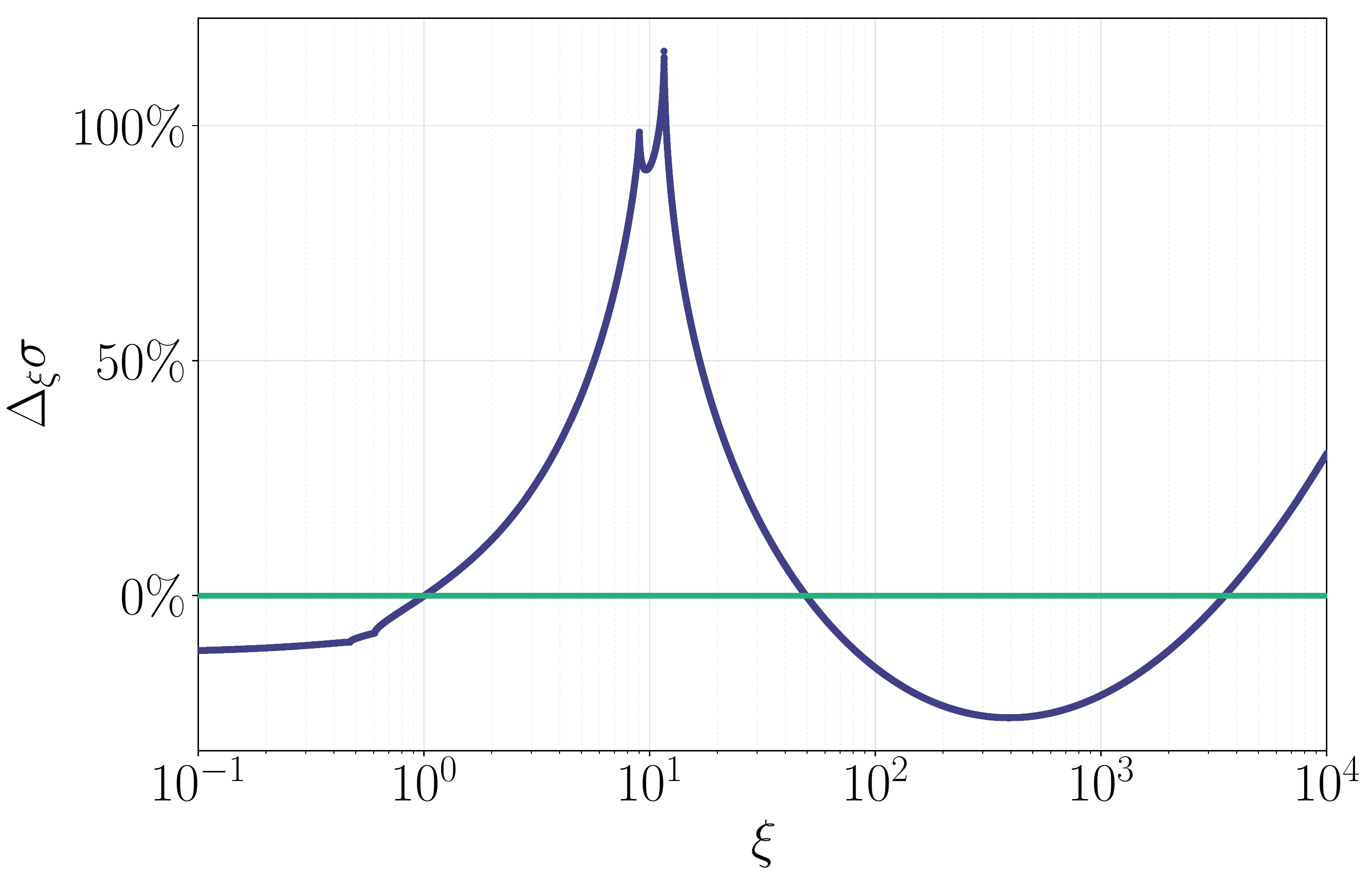}}
    \subfigure{\includegraphics[width=0.49\textwidth]{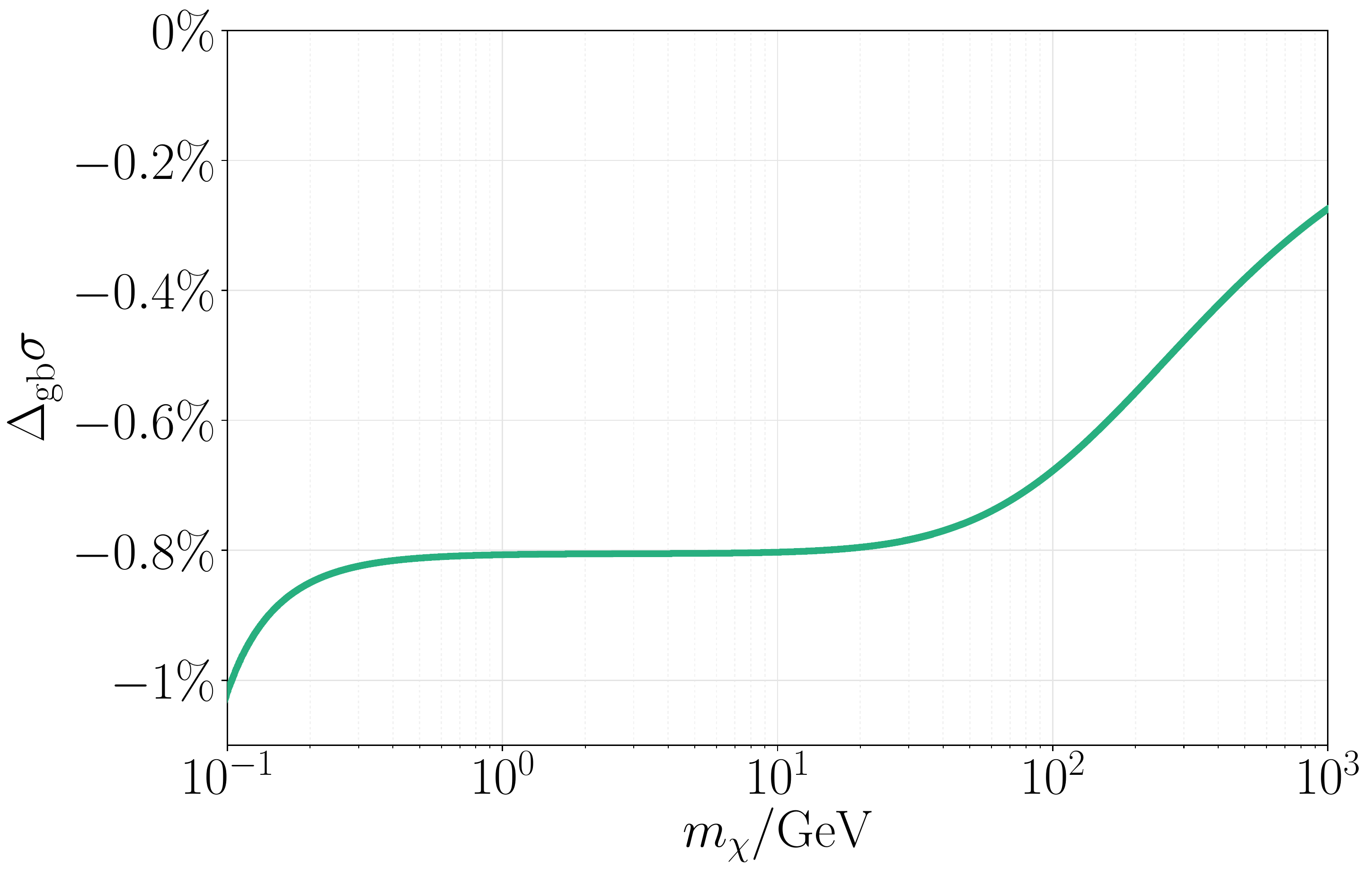}}
    \caption{Left: The relative change of the SI cross section as
      defined in \cref{eq::GaugeDep} is shown as a function of the
      gauge parameter $\xi$. The green line indicates the
      result calculated in our presented approach and the blue line
      corresponds to the approach including the additional gluon
      contributions discussed in \cref{sec::NLOGluonContributions}.; Right: The relative change of the SI
      cross section as defined in Eq.~(\ref{eq::TwoLoopBoxContributions}) as a function of the DM mass $\mX$. }
    \label{fig::GaugeNOGlu}
\end{figure}
We calculated all contributing diagrams in the general $R_{\xi}$ gauge
in order to be able to check for missing gauge
cancellations. As it turned out, our result is
completely gauge independent. For the proper cancellation of all gauge
dependencies the Goldstone triangle diagrams in
\cref{fig::boxTriangleTopologies} were crucial. They were needed to
properly cancel the gauge dependencies introduced in the vertex
corrections. These diagrams are often overseen in the
literature. However, the inclusion of the additional gluon
contributions introduces gauge dependencies which are not
cancelled. We define the relative change  
\begin{equation}
    \Delta_{\xi}\sigma \equiv \frac{\sigma-\left.\sigma\right|_{\xi=1}}{\left.\sigma\right|_{\xi=1}}\,,
    \label{eq::GaugeDep}
\end{equation}
where $\sigma$ indicates the SI cross section calculated with the
additional gluon contributions in the general $R_{\xi}$ gauge and in
the Feynman gauge ($\xi=1$), respectively. In the left plot of
\cref{fig::GaugeNOGlu} we show the results for the relative change as
a function of the gauge parameter $\xi$. The color scheme follows
  that of \cref{fig::CompApproaches}.  Again the benchmark scenario
presented in \cref{tab::BMP} is used to determine the SI cross
section. Obviously, when the additional gluon
  contributions are included, the variation of the gauge parameter
$\xi$ changes the SI cross section significantly preventing to make
reasonable predictions for the NLO SI cross section.  
Hence, not only the correct DM mass dependence is lost with the
inclusion of the additional gluon diagrams but also a
strong gauge dependence is introduced.  

Finally we will discuss the contribution of the third diagram in
\cref{fig:allDiagramsGluons}. In the right plot of \cref{fig::GaugeNOGlu} we show the relative difference 
\begin{equation}
    \Delta_{\text{gb}\sigma} \equiv \frac{\sigma-\left.\sigma\right|_{\text{nogb}}}{\left.\sigma\right|_{\text{nogb}}}\,,
    \label{eq::TwoLoopBoxContributions}
\end{equation}
where $\left.\sigma\right|_{\text{nogb}}$ is the SI cross section
calculated without the effective two-loop vertex (third diagram in
\cref{fig:allDiagramsGluons}) and $\sigma$ the SI cross section as
presented. 
We varied the DM mass $\mX$ while keeping the other input parameters
(same as in \cref{tab::BMP}) fixed. The most dominant effect of
the gluon boxes are obtained for small or large DM masses. Despite that, the overall impact given by the gluon boxes is in the sub-percentage region. Hence, not taking into account the gluon box
diagrams and thereby treating all diagrams with external gluons in
\cref{fig:allDiagramsGluons} consistently would not significantly
alter the overall result.

In addition to the phenomenological discussion of the SI cross section
of the \model~we implemented the model in {\tt BSMPT}
\cite{Basler:2018cwe,Basler:2020nrq} allowing us to check for a first-order 
electroweak phase transition in the early universe. For simplicity we
force the vacuum expectation value of the DM field component $\X$ in
\cref{eq:vevsExpansion} to be equal to zero at all temperatures while
determining the global NLO minimum at finite temperature. This way we
ensure the stability of the DM candidate and its DM nature. 
All parameter points above the neutrino floor provide an NLO stable
vacuum in the sense that the vacuum ground state of the one-loop
effective potential (at zero temperature) is the same as the
tree-level ground state. However, all parameter points provide a weak
phase transition $v_c/T_c<1$, where $v_c$ is the $SU(2)$ VEV ($v$) at
the critical temperature $T_c$. The critical temperature is defined as
the temperature where the one-loop effective potential has two
degenerate minima. A more involved study in the phase structure at
finite temperature of the \model~might enable a strong first order
electroweak phase transition. For instance, allowing the DM field
component to evolve a non-zero VEV at finite temperature leads to 
interesting phenomenological consequences. These studies are left for
future work.

\section{Conclusions}
\label{sec:Conc}

In this work we have calculated the NLO corrections to the spin
independent scattering cross section of a scalar DM particle off a nucleon in a Pseudo Nambu-Goldstone DM
model. This model has a scalar potential invariant under a global
$U(1)$ symmetry
softly broken such that a pseudo Nambu-Goldstone boson originates from the broken symmetry. The cross section was first shown to be proportional
 to the Dark Matter velocity in Ref.~\cite{PhysRevLett.119.191801}. Therefore there was the need to perform the calculation 
 at NLO. There were two independent calculations that appeared very close in time~\cite{Azevedo:2018exj, Ishiwata:2018sdi}. 
 
The first calculation~\cite{Azevedo:2018exj} was performed by
considering from the effective Lagrangian 
 \begin{equation}
    \mathcal{L}_{eff} = \sum_q C_S^q \mathcal{O}^q_S + C^g_S \mathcal{O}^g_S + \sum_{q} C_T^q \mathcal{O}^q_T\,,
\end{equation}
only the first term  $\sum_q C_S^q\mathcal{O}^q_S$. Instead of nuclear matrix elements for the proton an effective
Higgs-nucleon coupling was used. Because in this case the one loop result for the Wilson coefficient is independent
of the quark masses, it factorises, and it turns out that the Higgs-nucleon effective coupling is the sum of the nuclear
matrix elements. This calculation reproduces the correct dependence of the cross section in the limit of vanishing
Dark Matter and is at least 90\% of the total cross section
depending on the parameter points. Hence, relative 
to this work we have now included the terms $C^g_S \mathcal{O}^g_S + \sum_{q} C_T^q \mathcal{O}^q_T$.  
 
In the second calculation that appeared in Ref.~\cite{Ishiwata:2018sdi} all terms in the effective Lagrangian
were used. As previously discussed, the problem in this calculation resides in the gluon diagram contributions
(specifically the first two diagrams in~\cref{fig:allDiagramsGluons}). The 2-loop diagram is not effectively calculated and instead an 
approximation is performed such that a proper matching between the heavy quarks and the gluon operators 
cannot be performed. The only way to solve the problem would be to perform the complete 2-loop
calculation. The approximation leads therefore to the fact that the Goldstone nature of the DM candidate is not recovered
and that the these gluon contributions present a strong gauge dependence contrary to the rest of the calculation. 
Therefore these additional gluon contributions should be dropped
unless the full two-loop calculation is performed.

It is also worth mentioning that although we have used a different renormalisation scheme than the ones from the two previous calculations
our results show a very similar behaviour when the different contributions are compared. Finally, we showed that with the present constraints most of the allowed points are below the neutrino 
floor and only experiments in the far future will be able to probe
them.  

\section*{Acknowledgments}
We are thankful to Da~Huang and M.~Spira for fruitful and clarifying discussions.
R.S. is supported by the Portuguese Foundation for Science and Technology (FCT), under contracts
UIDB/00618/2020, UIDP/00618/2020, PTDC/FIS-PAR/31000/2017
and CERN/FISPAR/0002/ 2017, and by National Science Centre, Poland (NCN) HARMONIA project, contract
UMO-2015/18/M/S. The work of MM is supported by the
  BMBF-Project 05H18VKCC1, project number 05H2018.
\vspace*{0.5cm}

\bibliographystyle{bib-style}
\bibliography{dddm.bib}
\end{document}